\documentclass[11pt,a4paper]{article}
\usepackage{amsmath}
\usepackage{amssymb}
\usepackage{amsthm}
\usepackage{epsfig}
\usepackage{times}
\usepackage{wrapfig}
\usepackage{rotating}
\usepackage[numbers,sort&compress]{natbib}

\newcommand\restr[2]{{
  \left.\kern-\nulldelimiterspace 
  #1 
  \vphantom{\big|} 
  \right|_{#2} 
  }}

\usepackage[hang,stable]{footmisc}
\newcommand{\Sec}{\mathit{Sec}}
\newcommand{\MS}{m_{\scriptscriptstyle\mathrm{MS}}}
\newcommand{\reals}{\mathbb{R}}
\newcommand{\Identity}{\mathrm{id}}

\begin{document}
\ifx\href\undefined\else\hypersetup{linktocpage=true}\fi

\title{Does cosmological expansion affect local physics?}

\author{Domenico Giulini            \\
        Institute for Theoretical Physics and
        Riemann Center for Geometry and Physics\\
        Leibniz University Hannover \\
        Appelstrasse 2, D-30167 Hannover, Germany\\
        Center for Applied Space Technology and Microgravity\\
        University of Bremen\\
        Am Fallturm, D-28359 Bremen, Germany}

\date{}

\maketitle

\begin{abstract}
\noindent
In this contribution I wish to address the question 
whether, and how, the global cosmological expansion 
influences local physics, like particle orbits and 
black hole geometries. Regarding the former I 
argue that a pseudo Newtonian picture can be quite 
accurate if ``expansion'' is taken to be an attribute 
of the inertial structure rather than of ``space'' in 
some substantivalist sense. This contradicts the 
often-heard suggestion to imagine cosmological 
expansion as that of ``space itself''.  Regarding
isolated objects in full General Relativity, like black 
holes, I emphasise the need for proper geometric 
characterisations in order to meaningfully compare them 
in different spacetimes, like static and expanding ones. 
Examples are discussed in some detail to clearly map out 
the problems.

\end{abstract}

\begin{footnotesize}
\setcounter{tocdepth}{3}
\tableofcontents
\end{footnotesize}

\section{Introduction}
One of the most stunning statements in modern cosmology---%
i.e. cosmology after Einstein's seminal paper of 1917 
(\cite{Einstein:CP}), Vol.\,6, Doc.\,43, p.\,541-552)---
is that ``the Universe is expanding'', at least on average.
This provokes the question of what it is that expands, 
i.e., what object or structure is ``expansion'' really an 
attribute of? This is the question I wish to address in 
this contribution.  But before outlining its structure, 
let me briefly recall some historical background. 

As a theoretical possibility within the framework of 
General Relativity (henceforth abbreviated GR), global cosmological 
expansion was first conceived by Alexander Friedmann (1888-1925) 
in his 1922 paper~\cite{Friedmann:1922} and slightly later 
also in his more popular book ``The World as Space and Time'' 
of 1923, of which a German translation is available~\cite{Friedmann:DieWeltAlsRaumUndZeit}. 
However, the ``discovery'' of 
the Expanding Universe is nowadays attributed to Georges 
Lema\^{\i}tre (1894-1966) who was the first to use it as a 
possible explanation for the observed redshifts (mostly 
immediately interpreted as due to recession velocities) in 
the optical spectra of ``nebulae'' by Vesto Slipher (1875-1969).%
\footnote{The English translation of the originally french 
title of Lema\^{\i}tre's 1927 paper is: ``A homogeneous universe
of constant mass and increasing radius, accounting for the 
radial velocity of extragalactic nebulae''.}

\begin{figure}
\begin{center}
\includegraphics[width=0.9\linewidth]{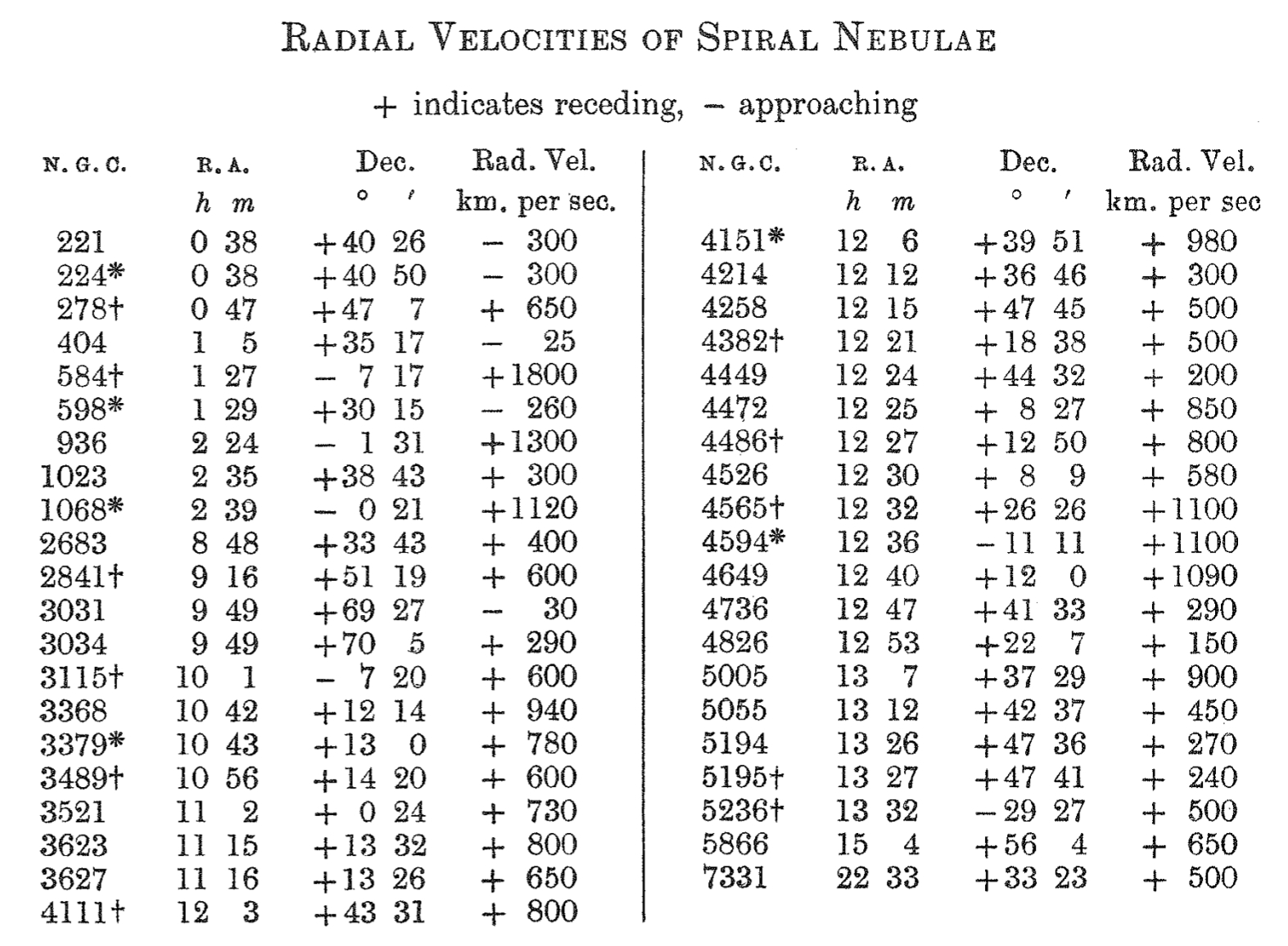}
\end{center}
\caption{\label{fig:EddingtonTable}\footnotesize
Table taken from Eddington's \emph{The Mathematical Theory of 
Relativity}. N.G.C. refers to the ``New General Catalogue'' as
published by John Louis Emil Dreyer in the Memoirs of the Royal 
Astronomical Society, Vol.\,49, in 1888. (The ``New'' refers to 
the fact that it succeeded John Herschel's ``General Catalogue 
of Nebulae'' of 1864.) This table was prepared for the second 
edition of Eddington's book by V.\,Slipher, containing, as of 
February 1922, a complete list of measured red shifts (velocities).
Entries marked with $*$ are said to be confirmed by others, 
those marked by $\dagger$ are said not to be as accurate as 
others. Of the five entries with negative radial velocities 
those approaching fastest are the Andromeda Galaxy NGC\,224 
(M\,31), its elliptic dwarf satellite galaxy NGC\,221 (M32),
and the Triangulum Galaxy NGC\,598 (M\,33), all of which are 
well within the Local Group. Of the other two, NGC\,3031 (M81) 
is Bode's Galaxy, whose modern (1993) distance determination 
gives 3,7\,Mpc. The other one, NGC\,404 (not listed in Messier's 
catalogue) is also known as ``Mirach's Ghost'' for it is hard to 
observe being close to the  second magnitude star Beta 
Andromedae. Up to ten years ago its distance was very uncertain. 
Since 2001, and with independent confirmations in the subsequent 
years, it has been determined to be at 3.1-3.3\,Mpc. 
The largest positive redshift in the list is shown by 
NGC\,584, which is $z\approx 6\times 10^{-3}$ corresponding to 
a recession velocity of 1800 km/s. This value is confirmed 
by modern measurements (NASA/IPAC Extragalactic Database). 
This is an elliptical galaxy in the constellation Cetus, 
the modern distance estimate of which is 23.4 according to 
the OBEY survey (Observations of Bright Ellipticals at Yale).}
\end{figure}

Slipher's results were brought to the attention of others mainly 
by Arthur S. Eddington (1882-1944), who included a list of 41 
radial velocities of spiral nebulae in the 2nd edition of 
his book \emph{The Mathematical Theory of Relativity} of 
1924. The list, provided by Slipher, is shown in 
Figure\,\ref{fig:EddingtonTable} and contains 41 ``nebulae'' 
(galaxies) in a distance range (modern values) roughly 
between 0.6 and 25 Mpc.\footnote{%
Mpc:=Megaparsec=$10^6\,\mathrm{parsec}$. 
$1\,\mathrm{parsec}\equiv 1\,\mathrm{pc}\approx 3.26\,
\mathrm{ly}\approx 3.1\,\times 10^{16}\,\mathrm{m}$; 
note $\mathrm{ly}\equiv$ lightyear.} Only five nebulae 
in Slipher's list show blueshifts, three of which are 
members of the local group and hence less than a Mpc away.
Roughly speaking, according to this table and modern 
(independent!) distance estimates, the dominance of 
recession sets in at about 10\,Mpc. However, this table 
gives redshifts only up to $z\approx 10^{-2}$ and neglects 
``southern nebulae'', as Eddington regretfully remarks. 

\begin{figure}
\begin{center}
\includegraphics[width=0.8\linewidth]{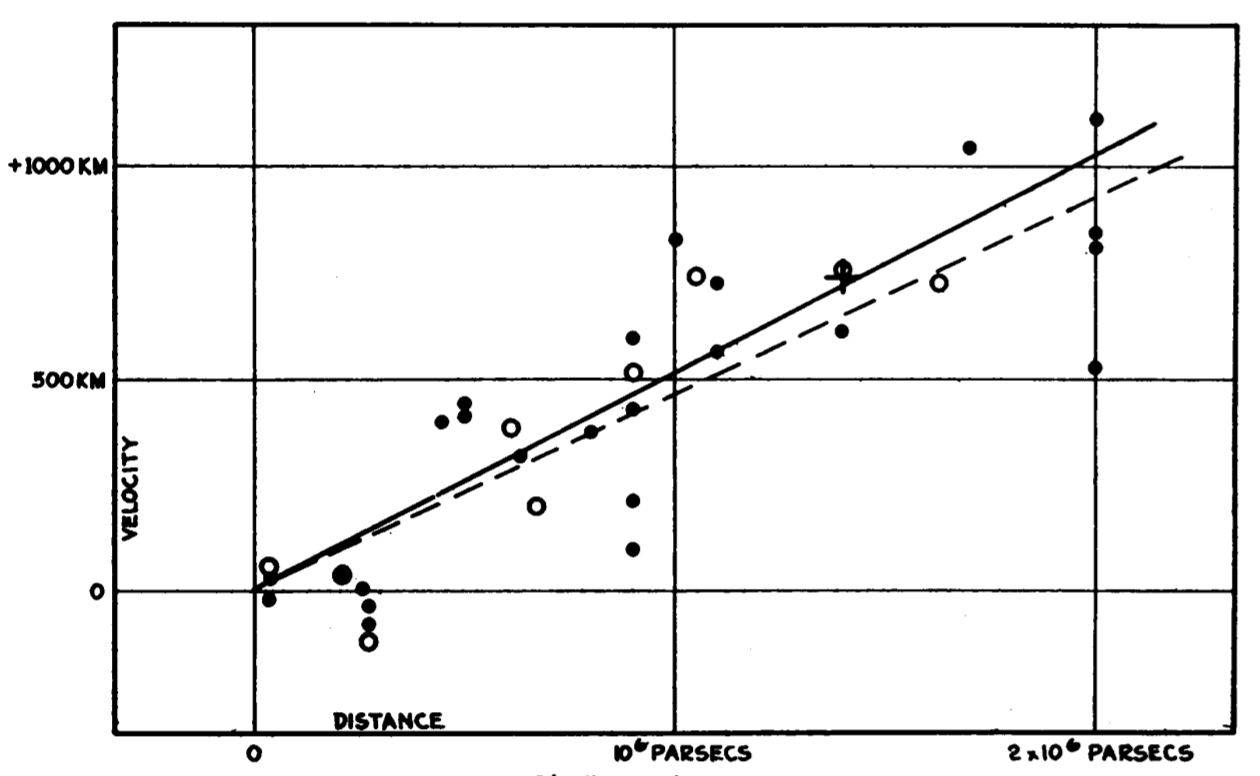}\\
\end{center}
\caption{\label{fig:HubblePlot1929}\footnotesize
Hubble's original plot (shown on p.\,172 of 
\cite{Hubble:1929}) of radial recession velocity, inferred 
from the actually measured redshifts by assuming Doppler's 
formula, versus distance. The latter is underestimated by 
a factor up to about~8.}
\end{figure}

It was Hubble who in his famous paper \cite{Hubble:1929} of 1929
explicitly suggested a \emph{linear} relation (to leading order) 
between between distances and redshifts/velocities, as 
shown in the well known plot from that paper, which we 
here reproduce in Figure\,\ref{fig:HubblePlot1929}.
Note that Hubble underestimated distances by a factor of 
about 8. For example, Hubble states the distance to the Virgo 
cluster as 2~Mpc, the modern value being 16.5~Mpc for our 
distance to its centre. 

\begin{figure}
\begin{center}
\includegraphics[width=0.9\linewidth]{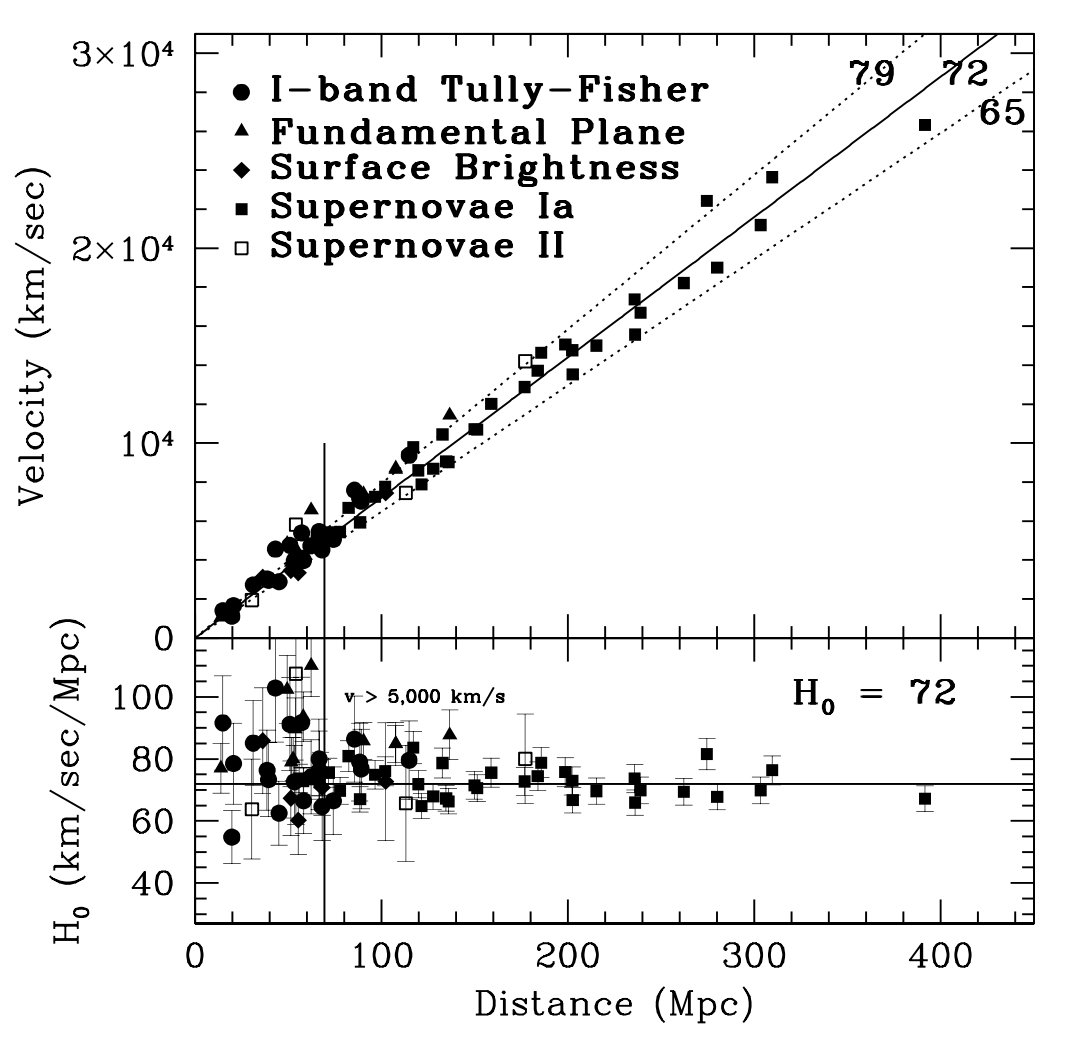}
\end{center}
\caption{\label{fig:HubblePlot2001}\footnotesize
Modern Hubble plot including the closer type I\,a 
supernovae up to distances of 400\,Mpc and redshifts 
of 0.1 (recession velocities of almost 10\% of the 
velocity of light). This plot is taken from the 
final analysis \cite{Freedman:2001} of the 
\emph{Hubble Space Telescope Key Project}. A slope of 
$H_0=72~\mathrm{km\cdot s^{-1}\cdot Mpc^{-1}}$ is shown 
flanked by $\pm 10\%$ lines. The plot below shows the 
apparent variation of $H_0$ with distance.}
\end{figure}

Modern Hubble plots include type\,Ia supernovae as standard 
candles. Figure\,\ref{fig:HubblePlot2001} shows a plot from
the final 2001 publication \cite{Freedman:2001} of the 
\emph{Hubble Key Project}, which includes closer ($z\lesssim 0.1$)
Ia supernovae calibrated against Cepheids. Note that this 
plot already extends in distance scale Hubble's original 
one (Figure\,\ref{fig:HubblePlot1929}) by a factor of 
about 25, and that the supernovae investigated by 
\emph{Supernova Cosmology Project} reach up to redshifts of 
about $z=1$, hence extending this plot by another factor 
of 10. Deviations from a linear Hubble law in the sense 
of an accelerating expansion are seen roughly above 
$z=0.4$.
See, e.g., \cite{Ellis:1989,Nussbaumer.Bieri:ExpandingUniverse}
for more on the intriguing  history of the ``expanding
universe''.   

At first sight, the above statement concerning the Universe 
being in a state of expansion is ambiguous and hardly understandable. 
It cannot even be taken face value unless we have a good idea of 
what ``Universe'' refers to. But this can be said precisely 
within the limits of relativistic \emph{cosmological models} about
which I will make some general remarks in the next section. 

In this contribution I will rather focus on the follow-up question of 
how to characterise structures that do and structures that do not 
participate in the expansion. For this I will first consider 
orbits of ``test'' masses (structureless masses of arbitrarily small
spatial extent whose own gravitational field is negligible) in 
the gravitational field of a central mass, the whole system being 
embedded in an expanding universe. In section\,2 I will employ  
a simple pseudo-Newtonian model for the dynamics of point particles 
in expanding universes. In this model the absolute simultaneity 
structure as well as the geometry for space and time measurements 
will be retained from Newtonian physics, but the inertial 
structure is changed so as to let the cosmological expansion 
correspond to force-free (inertial) motion. This will be achieved 
by adding an additional term to Newton's second law, in full 
analogy to the procedure one applies in other cases when 
rewriting Newton's second law relative to non-inertial reference 
frames.  This model, the intuitive form of which will be 
later justified in the context of GR, will lead us to reasonable first 
estimates for those systems that are themselves reasonably well described 
by Newtonian physics (i.e. excluding things like black holes). 
In doing this I will stress that the thing that undergoes expansion, 
either accelerated or decelerated, is the inertial structure and not some kind of ``space'' in the sense of substantivalism. The latter notion  would suggest the emergence 
of frictional or viscous forces on any body moving relative to 
that kind of ``space''. But, as we will also see in section\,2, 
this is not the right picture even though it is often stated 
(in particular in popular accounts) that ``it is space itself 
that expands''. Section\,3 discusses cosmological models in GR 
proper and justifies the pseudo-Newtonian approach regarding 
test-mass orbits of section\,2 from first principles in GR. It also 
comments on another effect that cosmological expansion has on 
the mapping of trajectories. This effect is more of a kinematical 
nature and arises from the fact that the notions of 
\emph{simultaneity} and \emph{instantaneous distance}, as defined 
by the geometry in standard cosmological models, are not identical 
with the corresponding notions using the exchange of light signals 
(Einstein simultaneity). 
Section\,4 discusses black holes in expanding universes, 
which are outside the realm of applicability of the 
pseudo-Newtonian picture. Rather, here we 
have to employ proper geometric techniques from GR in order to be
sure to characterise the physical situations independently 
of the coordinates used. Known exact solutions representing 
spherically symmetric black holes in expanding universes are 
discussed with an attempt to meaningfully characterise 
the impact of expansion. Finally, generalisations of a specific 
class of solutions are discussed along the lines of \cite{Carrera.Giulini:2010b}.  

A standard picture for global expansion is that of a 
rubber balloon being gradually filled with air; see, e.g., 
Figure\,27.2 in \cite{Misner.Thorne.Wheeler:Gravitation}. 
In such a picture the ``Universe'' is identified with the 
rubber sheet of a balloon. The two dimensional sheet is 
meant to represent three dimensional space. Points in 
real space off that sheet are simply not part of the model
and do not represent anything real. On that sheet we paint 
little circular discs and also glue some coins of about the 
same size. The painted elements of the rubber material 
continue to expand unhindered from each other, but underneath 
the coins the glue holds them rigidly in positions of 
unchanging mutual distances. We ask: which physical structures 
in the real world are meant to correspond to the painted 
and which to the glued coins? What sort of physical 
interactions can act like the glue in this picture? 

Note that in the pseudo-Newtonian discussion we pretend a 
clear split between space and time and that the 
``Universe'' at an ``instant'' corresponds to three-dimensional 
space filled with all there is. It is clear that in 
relativistic cosmology this corresponds to more structure 
than just a spacetime (four-dimensional differentiable 
manifold with Lorentzian metric) that satisfies the coupled 
field equations for the gravitational (metric) and matter fields.
What structure one needs in order to be allowed to talk in a 
Newtonian fashion will we recalled below. Until then let us 
proceed unworried guided by Newtonian intuition.     

If for the moment we assume that cosmological expansion 
were active within our solar system, we might be tempted 
to suspect it to cause dynamical anomalies, like extra 
radial accelerations. Such an anomalous acceleration 
had indeed been found for the Pioneer 10 and 11 
spacecrafts~\cite{Anderson.etal:2002a,Markwardt:2002}. 
Its possible cosmological origin is suggested by its 
observed magnitude 
\begin{equation}
\label{eq:PioneerAcceleration}
\Delta a=(8.6\pm 1.34)\times 10^{-10}\,\mathrm{m\cdot s^{-2}}\,,
\end{equation}
which is close to the product of the currently measured 
Hubble constant $H_0$ with the velocity of light
\begin{equation}
\label{eq:HubbleTimesC}
H_0\cdot c\approx (74\,\mathrm{km\cdot s^{-1}\cdot Mpc^{-1}})\cdot
(3\times 10^5\,\mathrm{km\cdot s^{-1}})\approx 7\times 10^{-10}\,
\mathrm{m\cdot s^{-2}}\,.
\end{equation}

Now, even though the \emph{Pioneer anomaly}, as it was called, 
has most likely received a far more mundane explanations recently~%
\cite{BertolamiEtAl:2008,Rievers:Laemmerzahl:2011,SchlaeppiEtAl:2012}, 
and even though it always seemed hard to believe that such a 
connection should exist at all, it was not entirely easy to show 
such an impossibility within a scheme of controlled approximations.
Note also that the sign was contrary to intuition: Whereas an 
accelerated expansion (like the one we presently seem to undergo) 
would give rise to extra outward-pointing accelerations (see below),
the measured anomalous accelerations of the Pioneer spacecrafts 
pointed inwards, more or less towards the Sun-Earth system. 
\begin{figure}
\begin{minipage}[c]{0.35\linewidth}
\centering\includegraphics[width=1.0\linewidth]%
{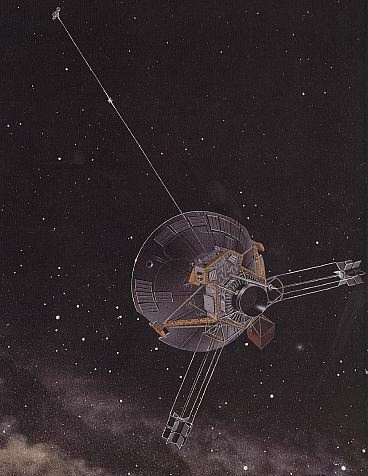}
\end{minipage}
\hfill
\begin{minipage}[c]{0.63\linewidth}
\centering\includegraphics[width=0.95\linewidth]%
{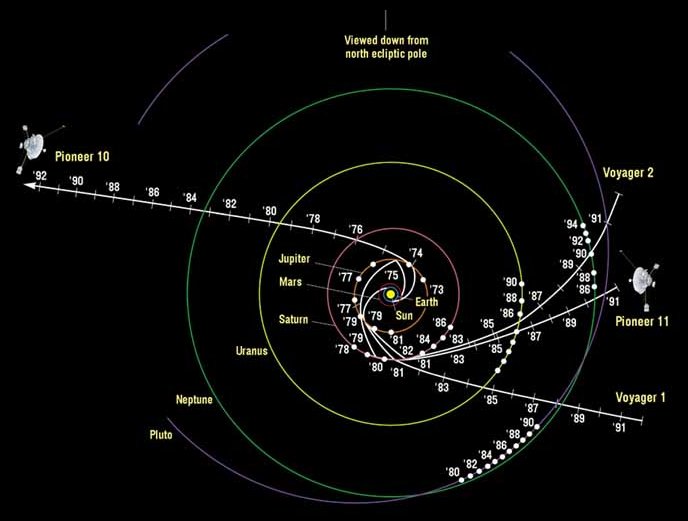}
\end{minipage}
\caption{\footnotesize
Left: Impression of the Pioneer satellite. 
Right: View from the north ecliptic pole onto the ecliptic showing 
the escape orbits of Pioneer\,10 and Pioneer\,11. The dots and 
numbers along the trajectories denote the positions and dates 
for each satellite.}
\end{figure}

\section{A Pseudo-Newtonian Picture}
\label{sec:PseudoNewtionanPic}
The rubber-sheet picture of cosmological expansion is useful 
to capture some \emph{kinematical} aspect. But, if 
na\"{\i}vely interpreted,  it is definitely misleading 
as far as the \emph{dynamical} aspects are concerned. 
The na\"{\i}ve and misleading interpretation is to think 
of expanding space as moving substance, composed of individuated 
points that can be assigned a local state of motion, like for 
a  fluid. This picture would suggest that a body in relative 
motion to the ``fluid'' will eventually be be dragged with it due to 
frictional or viscous forces. If this were true, the equations 
of motions should contain such force terms whose effect is to 
universally diminish velocities relative to the cosmic flow. 
But this is not the case, as has been discussed at length in \cite{Whiting-Alan:2004,Peacock:2006,Barnes.etal:2006}
and as we will see below. 
Rather than causing \emph{frictional forces}  proportional to 
the \emph{first} time-derivative of expansion parameter, the 
correct picture of the expansion's dynamical effect is to 
cause \emph{apparent forces} proportional to the 
\emph{second} time-derivative of the expansion parameter.
In other words, what cosmic expansion does is to change the 
inertial structure of space, so that coordinates based on 
metrically equidistant marks (as usually employed in 
the Newtonian equations of motion) become non inertial.  

\subsection{Changing the inertial structure}
\label{sec:ChangingInertialStructure}
A natural way to fit the general-relativistic concept of 
cosmological expansion into a pseudo-Newtonian framework 
is to recall that in GR the expanding structures 
move inertially (if represented by dust-like matter without 
pressure). This is transcribed into Newtonian language by 
keeping the absolute simultaneity structure and the 
geometries for space and time measurements (instantaneous 
space being still modelled by $\reals^3$ with its euclidean 
metric), but replacing the usual inertial structure such that 
inertial motion is now represented by wordlines of particular 
time-dependent instantaneous mutual separation (rather 
than being straight and parallel). This can be done by adding 
a suitable term to Newton's second law, just as it is done 
by rewriting this law so as to be valid in non inertial reference 
frames. We now give a simple derivation of this extra term.    

We require the local inertial frames to radially move apart 
according to Hubble's law: 
\begin{equation}
\label{eq:HubbleLaw-1}
\dot R(t)=H(t) R(t)\,.
\end{equation}
Here $R(t)$ is the instantaneous (with respect to some 
cosmological time) distance between two inertial frames 
and $H$ is the Hubble constant (the ``constant'' refers 
to it not depending on space). $H(t)$ is usually related 
to some cosmological scale parameter $a(t)$
via 
\begin{equation}
\label{eq:HubbleLaw-2}
H(t):=\dot{a}(t)/a(t)\,.
\end{equation}
This gives (suppressing arguments from now on) 
\begin{equation}
\label{eq:HubbleLaw-3}
\dot H=(\ddot a/a)-H^2=-H^2\bigl(1+q\bigr)\,,
\end{equation}
where 
\begin{equation}
\label{eq:HubbleLaw-4}
q:=-\ddot aa/{\dot a}^2
\end{equation}
is usually called the \emph{deceleration parameter}. 
Note that $H=\dot a/a$ and $\ddot a/a$ are decreasing 
with time $\propto t^{-1}$  and  $\propto t^{-2}$,
respectively, if $a(t)$ is a power-law and are constant 
in time if $a(t)$ grows exponentially. In any case, we assume 
that the typical timescales on which both quantities 
vary are much larger than the timescale over which 
the motions of the objects that we consider take 
place, so that we can replace $H$ and $q$ by their 
current values, indicated by a subscript $0$ (for 
$t=t_0=$ `now'). A recent evaluation \cite{KomatsuEtAl:2011},
based on the 7-year WMAP data, quotes as ``seven-year mean''
the following value for the Hubble constant\footnote{This result 
includes as prior the independent determination of the Hubble 
constant from a differential distance ladder \cite{RiessEtAl:2009}.}
\begin{equation}
\label{eq:WMAP-7-Data:Hubble}
H_0=(70.4\pm 2.5)\,\mathrm{km\cdot s^{-1}\cdot Mpc^{-1}}\,,
\end{equation}
and as values for the relative contributions of matter and 
(dark/vacuum-) energy to the overall gravitating energy
\begin{equation}
\label{eq:WMAP-7-Data:Omegas}
\Omega_m=0.274\pm 0.03\,,\quad
\Omega_\Lambda=0.727\pm 0.03\,.
\end{equation}
The deceleration parameter follows from this via the 
relation $q_0=\tfrac{1}{2}\Omega_m-\Omega_\Lambda$, which 
is an immediate consequence of the Friedmann equations. 
Hence we may take as currently best value  
\begin{equation}
\label{eq:DecParamValue}
q_0=-0.59\,.
\end{equation}

Proceeding in a pseudo-Newtonian fashion, we recall that 
the Newtonian force is proportional to the acceleration 
relative to the local inertial frames. The local inertial 
frames move radially according to \eqref{eq:HubbleLaw-1}.
Hence, using \eqref{eq:HubbleLaw-1} and \eqref{eq:HubbleLaw-3},
their outward acceleration is 
\begin{equation}
\label{eq:InertialSystAcceleration}
\ddot R=\dot HR+H\dot R=-HqR=\frac{\ddot a}{a}\,R\,.
\end{equation}

With respect to a given inertial frame, which we may 
choose as origin of $\reals^3$, the other inertial 
systems move with instantaneous velocity and acceleration
\begin{equation}
\begin{split}
 \dot{\vec x}&=H\vec x\,,\\
\ddot{\vec x}&=-qH^2\vec x=\frac{\ddot a}{a}\vec x\,,
\end{split}
\end{equation}
which is independent of the chosen origin.\footnote{%
Let $\vec v(t,\vec x)$ be an arbitrary velocity field 
in $\reals^3$, say of a fluid.  An observer who at time $t$ 
and position $\vec x$ is co-moving with the fluid sees the 
velocity distribution $\vec h\mapsto \vec w_{\vec x}(t,\vec h):=
\vec v(t,\vec h+\vec x)-\vec v(t,\vec x)$. This is independent 
of the observers location $\vec x$ if and only if 
$\vec v(t,\vec x)$ is an affine function of $\vec x$, i.e. 
$\vec v(t,\vec x)=A(t)\cdot\vec x+\vec a(t)$ for some 
matrix-valued function $A$ and vector-valued function 
$\vec a$ of time. Sufficiency is obvious and necessity follows 
since $\vec w_{\vec x}(t,\vec h)=\vec w_{\vec 0}(t,\vec h)$ can 
be simply rewritten into $\vec w_{\vec 0}(t,\vec h+\vec x)=
\vec w_{\vec 0}(t,\vec h)+\vec w_{\vec 0}(t,\vec x)$, showing 
that $\vec w_{\vec 0}$ must be linear in its second argument. 
The result now follows from 
$\vec v(t,\vec x)=\vec w_{\vec 0}(t,\vec x)+\vec v(t,\vec 0)$.
}

This means that the Newtonian equations of motion of a test 
particle in an expanding universe are obtained from the usual 
one by replacing  
\begin{equation}
\label{eq:ModNewEq-1}
\ddot{\vec x}\mapsto \ddot{\vec x}-(\ddot a/a)\,\vec x \,.
\end{equation}
We will later discuss how this replacement finds a simple 
explanation in GR. With it, the modified 
Newtonian equation then reads
\begin{equation}
\label{eq:ModNewEq-2}
m\bigl(\ddot{\vec x}-(\ddot a/a)\,\vec x\bigr) =\vec F\,.
\end{equation}
We end this subsection by making the simple general observation 
that \eqref{eq:ModNewEq-2} involves only the \emph{second} 
time derivative of $a(t)$. Hence the sign of $\dot a$ does not 
matter, as would be the case if the impact of cosmological 
expansion would be analogous to some sort of viscous or 
frictional force due to the relative motion against a 
substantivalist's ``space''. A given positive $\ddot a/a$ can 
either be caused by a universe in a state of accelerated 
expansion or decelerated contraction. Likewise, a given negative 
$\ddot a/a$ can either be caused by a universe of decelerated 
expansion or accelerated contraction.

\subsection{Inertial motion}
\label{sec:Inertial motion}
In this subsection we briefly discuss solutions to 
\eqref{eq:ModNewEq-2} for $\vec F=0$ and special expansion 
laws $a(t)$. This is meant to illustrate the last point 
of the previous subsection. We start with the so-called 
matter-dominated universe in a state of decelerated expansion, 
given by $a(t)\propto t^{2/3}$. This gives   
\begin{equation}
\label{eq:MatterDomination}
\ddot a(t)/a(t)=-\frac{2}{9}\cdot t^{-2}\,.
\end{equation}
Integrating \eqref{eq:ModNewEq-2} for the initial conditions 
$\vec x(1)=(R,0,0)$ and $\dot{\vec x}(1)=\vec 0$, corresponding 
to a body that at time $t=1$ is released at distance $R$ on the 
$x$-axis with zero velocity, we obtain the following solution 
for the $x$ coordinate (the $y$ and $z$ coordinates stay zero)
\begin{equation}
\label{eq:MatterDomination-Solution}
x(t)=R\bigl(2t^{1/3}-t^{2/3}\bigr)\,.
\end{equation}
So even though ``space expands'', the particle first 
approaches the origin $x=0$, hits it at $t=8$ (i.e. 
after seven further unfoldings of the ``world age''), and 
then recedes from it along the negative $x$-axis in an 
asymptotically co-moving fashion. That despite expansion 
the particle first starts to approach $x=0$, rather than 
recede from it, as a frictional force would imply, is not 
too surprising in view of the fact that the initial condition
$\dot x(1)=0$ means that, \emph{relative to the inertial 
frame at the initial position $x(1)=R$}, the particle is 
moving with velocity $\dot a/a$ \emph{towards} the origin. 
The fact that the particle asymptotically approaches a 
co-moving state is not universally implied by the equations 
of motion, as one can see from the remark that \eqref{eq:MatterDomination}, 
and hence \eqref{eq:MatterDomination-Solution}, are the same for 
$a(t)\propto t^{1/3}$, in which case the particle overshoots 
the cosmic expansion for large times.\footnote{Quite generally,
an FLRW universe with perfect-fluid matter and equation of state 
$p=w\rho c^2$, where $w>-1$, expands like $a(t)\propto t^n$ 
with $n=2/3(w+1)$, so that $n=1/3$ corresponds to $w=1$, the 
extreme positive-pressure case in view of energy dominance.}

Similarly, for an exponential scale factor $a(t)\propto\exp(\lambda t)$, 
where $t$ now ranges over the full real axis, we have  

\begin{equation}
\label{eq:ExpExpansion}
\ddot a(t)/a(t)=\lambda^2\,.
\end{equation}
Picking the corresponding initial data $x(0)=R$ and 
$\dot x(0)=0$ for \eqref{eq:ModNewEq-2} we now get  
\begin{equation}
\label{eq:ExpExpansion-Solution}
x(t)=R\cosh(\lambda t)\,.
\end{equation}
Note that this is insensitive of the sign of $\lambda$. 
For $\lambda>0$ this might confirm the na\"{\i}ve expectation, 
since the particle immediately starts to recede from the 
origin and asymptotically approaches a co-moving state.
But this is certainly not true for a contracting universe 
where $\lambda<0$.

\subsection{Coulomb-like potential}
\label{sec:Coulomb-like-potential}
Let us now apply \eqref{eq:ModNewEq-2} to the case where 
$\vec F$ is a time independent radial force $\propto 1/r^2$.
To keep matters simple we will assume $\ddot a/a$ to be
constant. This is exactly true if $a(t)\propto\exp(\lambda t)$,
or approximately for motions during timescales in which 
$\ddot a/a$ changes very little. The second term in 
\eqref{eq:ModNewEq-2} then acts like a time-independent radial 
pseudo-force, which is outward pointing for either accelerated 
expansion or decelerated contraction, and inward pointing for 
decelerated expansion or accelerated contraction. The current 
value of this relative acceleration of inertial frames per 
unit of separation distance can be inferred from the data given 
above. Using $\ddot a_0/a_0=-q_0H_0^2$ and the values from 
\eqref{eq:WMAP-7-Data:Hubble} and \eqref{eq:DecParamValue} gives  
\begin{equation}
\label{eq:RelAccValue}
A:=\frac{{\ddot a}_0}{a_0}=-q_0H_0^2
\approx 10^{-13}\cdot\mathrm{m\cdot s^{-2}\cdot Mpc^{-1}}\,,
\end{equation}
which looks very small indeed! Here we introduced the letter $A$ 
as abbreviation for later notational convenience. 

We now put a time-independent radial $1/r^2$ force on 
the right-hand side of \eqref{eq:ModNewEq-2},
\begin{equation}
\label{eq:RadialForce}
\vec F=-m\vec\nabla\left(-\frac{C}{r}\right)\,,
\end{equation}
where $C$ is a constant which we assume to be positive since we 
are only interested in attractive central forces. 

As usual, time independence and rotational symmetry allow us to 
infer conservation laws for energy and angular momentum. The latter 
implies in particular that the motion stays in a plane, which we 
coordinatise by planar polar coordinates $(r,\varphi)$. Denoting 
by $E$ the energy per unit mass and by $L$ the modulus of angular 
momentum per unit mass, the conservation laws are  
\begin{equation}
\label{eq:ModNewEqRadialField-a}
\tfrac{1}{2}{\dot r}^2+U(r)=E\,,\qquad r^2\dot\varphi=L\,,
\end{equation}
where 
\begin{equation}
\label{eq:ModNewEqRadialField-b}
U(r)=\frac{L^2}{2r^2}-\frac{C}{r}-\frac{A}{2}\, r^2\,.
\end{equation}
This differs from the usual expression by the last term. 
In order to gain a first estimate of its impact consider 
the case $A>0$, e.g. accelerated expansion. In this case 
it leads to an additional local maximum to the right of 
the minimum describing stable circular orbits. For values 
of $r$ grater than that of the additional maximum the 
system is doomed to expand forever by following the cosmic 
expansion. The critical radius at which cosmic expansion 
just balances the attraction for a particle at rest $(L=0)$ is   
\begin{equation}
\label{eq:CriticalRadius-1}
r_c=\left[\frac{C}{A}\right]^{1/3}\,.
\end{equation}
As we will see in more detail below, this critical radius is an 
approximate upper bound for radii of stable circular orbits. 
Let us estimate its size. If attraction is due to the gravitational 
force of a central mass $M$ upon a test mass $m$, or the Coulomb
force of a central charge $Ze$ upon an elementary charge $e$
(also of mass $m$), we have 
\begin{equation}
\label{eq:VariousForces}
C=
\begin{cases}
GM & \text{gravitational field}\\
\frac{Qe}{4\pi\varepsilon_0m} &\text{electric field}\,.
\end{cases}
\end{equation}
The corresponding critical radii \eqref{eq:CriticalRadius-1}  
can then be expressed as weighted geometric mean of the 
Hubble radius 
\begin{equation}
\label{eq:HubbleRadius}
R_H:=\frac{c}{H_0}
=4.23\times 10^9\,\mathrm{Mpc}
=13.7\times 10^9\,\mathrm{ly}
=1.3 \times 10^{26}\,\mathrm{m} 
\end{equation}
with the Schwarzschild radius for the central mass $M$
(gravity case)
\begin{equation}
\label{eq:SchwarzschildRadius}
R_M:=\frac{2GM}{c^2}
\approx \left(\frac{M}{M_\odot}\right)\cdot 
3\times 10^3\,\mathrm{m}\,,
\end{equation}
or with the classical charge-radius\footnote{The classical 
charge-radius of a charge $Q$ with mass $m$ is defined 
to be the radius outside which the energy stored in 
the Coulomb field of charge $Q$ equals $mc^2$. 
Sometimes twice of that is called the ``classical 
charge radius''.} for the charge $Q$ with mass $m$
(electric case)
\begin{equation}
\label{eq:ChargeRadius}
R_Q:=\frac{Q^2}{8\pi\varepsilon_0mc^2c^2}
\approx \left(\frac{Q}{e}\right)^2\cdot 
1.4\times 10^{-15}\,\mathrm{m}\,.
\end{equation}
Explicitly, the simple expression for the critical 
radius in the gravitational case is 
\begin{equation}
\label{eq:CriticalRadius-gravity}
r^{(gr)}_c=
\bigl[-1/2q_0\bigr]^{1/3} \cdot\,\bigl[R_MR_H^2\bigr]^{1/3}
\approx \left(\frac{M}{M_\odot}\right)^{1/3}\ 352\,\mathrm{ly}\,,
\end{equation}
and in the electric case\footnote{$\mathrm{AU}\equiv$
astronomical unit $\approx 1.5\times 10^{11}\,\mathrm{m}$.} 
\begin{equation}
\label{eq:CriticalRadius-electric}
r^{(el)}_c=
\bigl[-2e/Qq_0\bigr]^{1/3}\cdot\,\bigl[R_Q\,R_H^2\bigr]^{1/3}
\approx \left(\frac{Q}{e}\right)^{1/3}\ 30\,\mathrm{AU}\,.
\end{equation}
The critical radii are the characteristic scales above which 
systems, which were bound if no expansion existed, disintegrate
as a result of accelerated cosmological expansion or decelerated 
contraction. 
  
The last equations \eqref{eq:CriticalRadius-electric} shows 
that disintegration of a Hydrogen atom is inevitable if the 
electron-proton distance is of the order of 30 astronomical 
units, that is about the semi-major axis of the Neptune orbit! 
Atoms, humans, and all things around us are essentially unaffected 
by cosmic expansion. For gravitating systems, \eqref{eq:CriticalRadius-gravity} 
sets the scale above which the solar system starts to 
disintegrate beyond 300 lightyears. Recall that the next star 
(Proxima Centauri) is about 4 lightyears away. If the 
central body is of about $10^{12}$  solar masses, like our 
Galaxy, expansion sets in above a scale above 3 million lightyears, 
that is 1.5 times farther than the distance to the Andromeda 
galaxy. Finally, if the consider a central mass of $10^{15}$ 
solar masses, like for the Virgo cluster, expansion sets in 
above 30 million lightyears or 10 megaparsecs. This is now, 
finally, smaller than the distance to other (smaller) clusters, 
like Fornax ($\approx$ 30 megaparsecs), and even larger than 
the distance to the local group ($\approx$ 20 megaparsecs). 
This leads one to roughly estimate the scale in our Universe 
above which gravitationally interacting systems start to follow 
the Hubble flow to be that of large galaxy clusters. Structures 
below that size are expected to stay bounded. This fits well 
with the observational status described in the introduction.
Note that the absence of any expansion below certain distances 
does not fit with the statement that ``space itself expands'', 
as there is space everywhere. 

\subsection{Existence of stable circular orbits}
Circular orbits are those of unchanging radius, i.e. $r(t)=r_*$.
Hence $\dot r(t)=0$ for all times $t$. It follows from 
\eqref{eq:ModNewEqRadialField-a} that that $U$ must have a 
stationary point at $r=r_*$:
\begin{equation}
\label{eq:SCO-1}
U'(r=r_*)=-\frac{L^2}{r_*^3}+\frac{C}{r_*^2}-Ar_*=0\,.
\end{equation}
If the circular orbit is to be stable, the stationary 
point must be a minimum:
\begin{equation}
\label{eq:SCO-2}
U''(r=r_*)=3\frac{L^2}{r_*^4}-2\frac{C}{r_*^3}-A>0\,.
\end{equation}
Adding $(3/r_*)$ times \eqref{eq:SCO-1} to \eqref{eq:SCO-2} 
gives
\begin{equation}
\label{eq:SCO-3}
\frac{C}{r_*^3}-4A>0\,.
\end{equation}
Since we restrict to $C>0$ this is automatically satisfied 
if $A<0$, i.e. for decelerated expansion or accelerated contraction.
However, for accelerated expansion or decelerated contraction
we get the non-trivial constraint
\begin{equation}
\label{eq:SCO-4}
r_*<4^{-1/3} r_c\approx 0.63\cdot r_c\,,
\end{equation}
where $r_c$ was defined in \eqref{eq:CriticalRadius-1}. 
This gives the precise upper bound on stable circular orbits. 
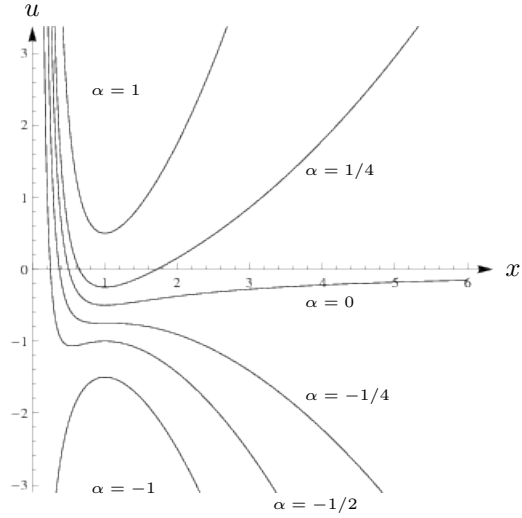
\begin{figure}
\begin{center}
\includegraphics[width=0.5\linewidth]{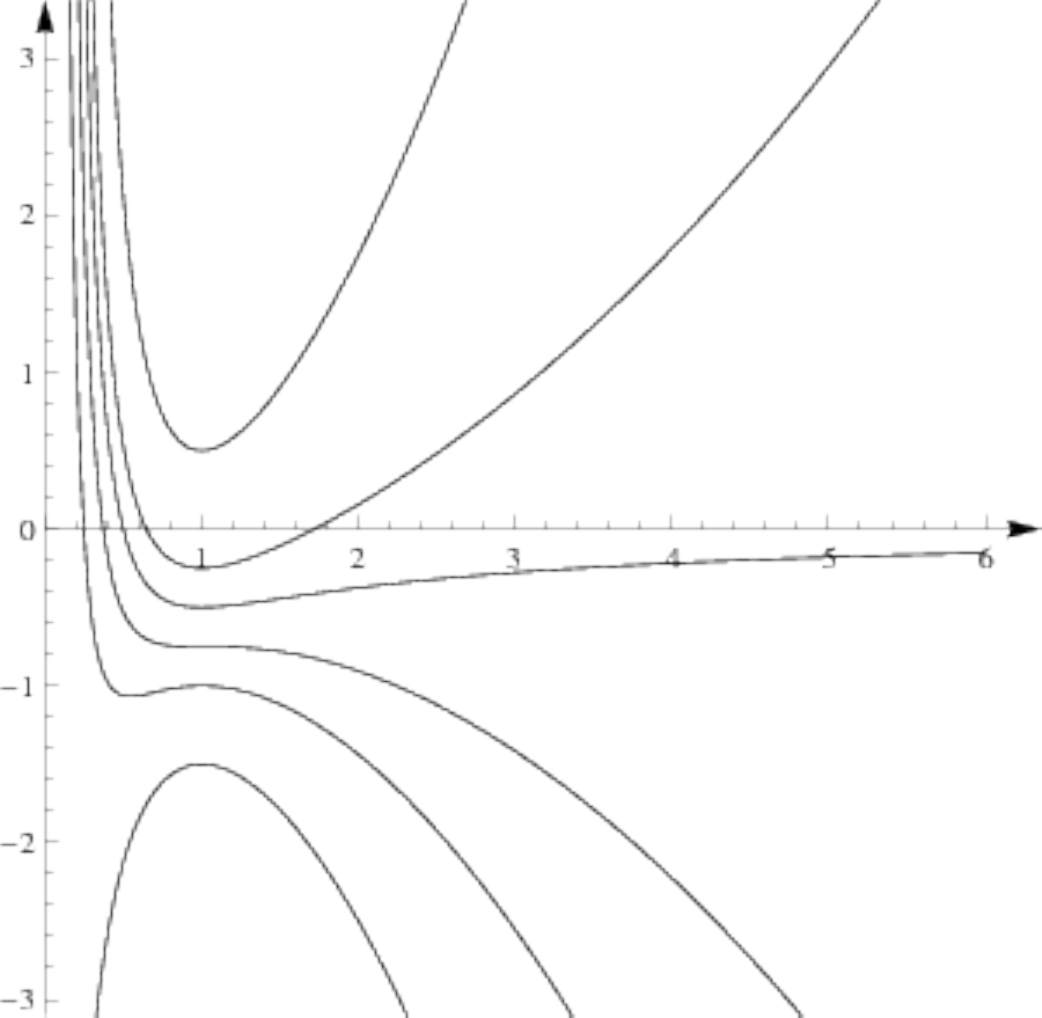}
\put(-150,150){\tiny $\alpha=1$}
\put(-70,120){\tiny $\alpha=1/4$}
\put(-70,70){\tiny $\alpha=0$}
\put(-70,35){\tiny $\alpha=-1/4$}
\put(-82,-6){\tiny $\alpha=-1/2$}
\put(-150,0){\tiny $\alpha=-1$}
\put(5,82){\small $x$}
\put(-175,180){\small $u$}
\end{center}
\caption{\label{fig:EffectivePotential}\footnotesize
This figure, taken from \cite{Carrera.Giulini:2010a}, 
shows various plots of the potential 
\eqref{eq:ModNewEqRadialField-b}
in the rescaled form \eqref{eq:ReducedPotential-1}. 
Circular orbits correspond to the extremum at $x=1$,
which are stable (minimum) for $\alpha<1/4$ and unstable 
$\alpha>1/4$.}
\end{figure}

The stability properties of the potential 
\eqref{eq:ModNewEqRadialField-b} are summarised in 
Figure\,\ref{fig:EffectivePotential}, where we plotted the 
rescaled potential $u:=\left(\frac{r_*}{C}\right)\cdot U$ 
as a function of $x:=\frac{r}{r_*}$. Here $r_*$ is the radius 
at which $U'(r_*)=0$, i.e. it satisfies \eqref{eq:SCO-1}.
We then have
\begin{equation}
\label{eq:ReducedPotential-1}
u(x)=\frac{1-\alpha}{2x^2}-\frac{1}{x}-\frac{\alpha}{2}x^2\,,  
\end{equation} 
with
\begin{equation}
\label{eq:ReducedPotential-2}
\alpha:=r_*^3\cdot\frac{A}{C}\,.
\end{equation} 
By construction $u(x)$ has an extremum at $x=1$.  
In Figure\,\ref{fig:EffectivePotential} we plotted \eqref{eq:ReducedPotential-1} for various values of
$\alpha\in[-1,1]$, with $-1\leq\alpha<0$ corresponding 
to the case of decelerated expansion or accelerated contraction
and  $0<\alpha\leq 10$ to accelerated expansion or decelerated 
contraction.  For increasing values of $\alpha$ the right 
slope of the graph comes down so as to turn the extremum at 
$x=1$, which is a minimum for $\alpha<1/4$, to a maximum for
$\alpha>1/4$. This corresponds to the transition from stable to
unstable circular orbits. 

The angular frequency $\omega_*$ of a circular orbit follows 
from the expression \eqref{eq:SCO-1} for $U'(r_*)=0$ if we 
set $L=r_*^2\omega_*$. We get 
\begin{equation}
\label{eq:KeplerThirdModified}
\omega_*=\omega_K\sqrt{1-r_*^3\frac{A}{C}}
\end{equation}
where 
\begin{equation}
\label{eq:KeplerThirdOriginal}
\omega_K:=\sqrt{\frac{C}{r_*^3}}
\end{equation}
is the `Keplerian orbital frequency of the unperturbed 
problem. Switching on $A$ changes Kepler's 3rd law from 
the familiar form $\omega_*=\omega_K$ to its modified form \eqref{eq:KeplerThirdModified}

If we compare circular orbits of a fixed radius 
$r=r_*$ of the unperturbed ($A=0$) with the perturbed problem,
\eqref{eq:KeplerThirdModified} tells us that the angular frequency 
is diminished if $A>0$ and enhanced if $A<0$. This clearly 
fits well with intuition since the additional force due to 
$A\not = 0$ is directed parallel ($A>0$) or opposite ($A<0$) 
the centrifugal force, so that in order to maintain 
equilibrium of forces we have to diminish or enhance the 
latter. If we adiabatically switched on cosmological expansion, 
the system would just readjust itself according to the 
modified Keplerian law \eqref{eq:KeplerThirdModified}. 
For small $A$ changes in the periods and radii are of the order 
$(r_*/r_c)^3$, where $r_c=(C/\vert A\vert)^{1/3}$, which are small
indeed (compare discussion above).

\section{The General-Relativistic Picture}
\label{sec:GR-Picture}
General relativistic models of spacetime differ from such models in 
Newtonian physics in several aspects. Very loosely speaking,
less structure is assumed in GR than is in Newtonian contexts. 
The word \emph{spacetime} in GR usually just refers to a tuple $(M,g)$, 
where $M$ is a smooth 4-dimensional manifold and $g$ is a 
(piecewise) smooth Lorentzian metric. Sometimes the word \emph{spacetime} 
is reserved to those tuples $(M,g)$ where $g$ obeys Einstein's 
equations with suitable matter sources, but here we need not be specific 
about that. The first thing we wish to stress -- following Hermann Weyl -- 
is that a \emph{cosmological model} comprises more structure than just a 
spacetime.

\subsection{On the notion of ``cosmological model'' in GR}
\label{sec:GR-CosmModel}
As already remarked in the introduction, we have to first 
understand what ``Universe'' refers to if we want to 
understand the statement concerning its alleged expansion. 
In order to answer this question in the context and geometric 
spirit of GR (i.e. in terms of geometric 
structures rather than special coordinate system), we recall 
that according to Weyl \cite{Weyl:RZM1991}
the definition of a \emph{cosmological model} comprises not 
only a spacetime $(M,g)$, but also a normalised timelike 
vector field $V$ on $M$. In GR a cosmological 
model thus comprises at least a triple $(M,g,V)$. The r\^ole of 
$V$ is to represent the flow of cosmological dust matter 
(sometimes referred to as ``privileged observers''). 
Already at this point we may speak of (local) expansion/contraction, 
namely if $V$ (locally) has positive/negative divergence. 
Weyl stresses \cite{Weyl:2009} that without the geometric 
structure supplied by, and the interpretation given to, 
$V$ one could not even attempt to calculate the cosmological 
redshifts, which is always to be thought of as taking place 
between systems moving along different flow lines of $V$.%
\footnote{Here we do not wish to enter into the discussion 
of \emph{Weyl's principle}, which further asserts that the 
flow lines of $V$ should be contained in a region of common
causal dependence, i.e. that no particle horizons should 
exist~\cite{Ehlers:2009}. See, e.g., \cite{Goenner:2001} and \cite{Rugh.Zinkernagel:2009} for discussions of partly 
alternative viewpoints on this principle.} 
Now, generally one would of course require that $g$ and $V$ 
be related by Einstein's equation.  For example, the energy 
momentum tensor of a perfect fluid (no friction, no heat 
conduction) is just 
\begin{equation}
\label{eq:DustEMT}
T=\rho V\otimes V +p\bigl(c^{-2}V\otimes V-g^{-1})\,,
\end{equation}
where $\rho$ is the fluid's mass density and $p$ is its pressure, 
both measured in the local rest frame. $V$ is the fluid's 
four-velocity vector field  (normalised to $g(V,V)=c^2$) and 
$g^{-1}$ is the ``inverse metric'' field (metric in the 
cotangent spaces).  The integrability condition of Einstein's 
equation, which says that $T$ must have vanishing covariant 
divergence with respect to $g$, then implies the relativistic 
Euler equations for $V$, which in the pressureless case, $p=0$,  
imply that $V$ is geodesic with respect to $g$. If, in addition, 
$V$ is hypersurface orthogonal, i.e. of vanishing vorticity, we 
may address the hypersurfaces orthogonal to $V$ as ``space''
(at least locally). Such a hypersurface may then be called 
an ``instant''. The ``Universe at an instant'' would consist 
of the instant and all metric and matter fields comprising 
Cauchy data. That ``the Universe (locally) expands'' usually  
just means that Cauchy surfaces (locally) increase their volume 
along the flow of $V$.  

Sometimes one just considers geodesic timelike vector fields 
$V$ on a given spacetime $(M,g)$ that solves Einstein's equation 
\emph{without} \eqref{eq:DustEMT} as source, that is, one 
neglects the back-reaction of the dust matter onto the geometry
of spacetime. Then it should not come as a surprise that the 
choice of $V$ is ambiguous. The most trivial example is perhaps 
flat Minkowski space, where, in standard coordinates, 
we could just take 
\begin{equation}
\label{eq:MinkVectorField-1}
V_1=\partial/\partial t
\end{equation}
which is even a geodesic Killing field. Another choice, defined 
in the upper wedge region $W:=\{(t,x,y,z)\in\reals^4\mid ct>r:=\sqrt{x^2+y^2+z^2}$, would be as follows: Instead of 
$(t,r)$ use coordinates $(T,\rho)$ in W, where 
\begin{equation}
\label{eq:NewCoord-1}
t=T\sqrt{1+\rho^2}\,,\quad
r=cT\rho  
\end{equation}
with inverse
\begin{equation}
\label{eq:NewCoord-2}
T=\sqrt{t^2-r^2/c^2}\,,\quad
\rho=\frac{r}{\sqrt{c^2t^2-r^2}}\,.  
\end{equation}
The hypersurfaces of constant $T$ are spacelike hyperboloids of 
constant timelike distance $T$ from the origin. Their intrinsic
metric is of constant negative curvature $T^{-2}$. They are 
orthogonal to the timelike geodesic vector field  
\begin{equation}
\label{eq:NewVectorField}
V_2=\frac{\partial}{\partial T}=\frac{1}{\sqrt{t^2-r^2/c^2}}
\left(
t\frac{\partial}{\partial t}
     +r\frac{\partial}{\partial r}
\right)\,.
\end{equation}
The Minkowski metric, restricted to $W$, can be written as 
\begin{equation}
\label{eq:MinkMetrik}
\begin{split}
\restr{g}{W}
&=-c^2\,dt^2+dr^2+r^2(d\theta^2+\sin^2\theta\,d\varphi^2)\\
&=-c^2\,dT^2+c^2T^2\left(\frac{d\rho^2}{1+\rho^2}
   +\rho^2(d\theta^2+\sin^2\theta\,d\varphi^2)\right)\,.
\end{split}
\end{equation}
This shows that the region $W$ of Minkowski spacetime can 
be written as an ``open'' (i.e. spatially negatively curved)  
FLRW model with scale function $a(T)=T^2$, which is usually 
called the  Milne model. Clearly $V_2$ is not Killing. 

A more interesting variety exists for the de\,Sitter 
spacetime, which is a solution $(M,g)$ to the vacuum Einstein 
equations with cosmological constant $\Lambda>0$. It can be 
represented as one-sheeted hyperboloid in five-dimensional 
Minkowski space:%
\footnote{The global structure of de\,Sitter's solution was 
revealed by Lanczos, Weyl, and, last not least, Felix Klein.
The contribution of Klein is often overlooked. See 
\cite{Roehle:2002} for a fair account.}   
\begin{equation}
\label{eq:deSitterHyperboloid}
M=\{(x^0,x^1,x^2,x^3,x^4)\in\reals^5\mid
-(x^0)^2+(x^1)^2+(x^2)^2+(x^3)^2+(x^4)^2=3/\Lambda\}
\end{equation}
and where $g$ is the metric induced from ambient Minkowski 
space. Various timelike and hypersurface orthogonal geodesic 
vector fields $V$ exist, either globally or on sub-domains of 
$M$, which allow to represent these domains as FLRW universes 
with positive, zero, or negative spatial curvature, expanding 
with $\cosh$, $\exp$ and $\sinh$ of time respectively. The 
corresponding homogeneous spacelike hypersurfaces are given 
by taking the intersection of $M$ with the following three 
families of spacelike, lightlike, and timelike hyperplanes 
in $\reals^5$: $x^0=\mathrm{const.}$, $x^0+x^1=\mathrm{const.}>0$, 
and $x^1=\mathrm{const.}>1$ with $x^0>0$.  

It is not necessary to write down explicit expressions for 
the various vector fields $V$. This can be done, e.g., from 
the expressions one obtains for the FLRW forms of the metric 
by setting $V=\partial/\partial T$, where $T$ is the FLRW 
time. Coordinate expressions of the FLRW forms corresponding 
to the slicings just mentioned are well known\footnote{See, 
e.g., http://en.wikipedia.org/wiki/De${}_-$Sitter${}_-$space}

This shall suffice to characterise the ambiguity in representing 
a given spacetime $(M,g)$ as a sequence of Universes, i.e. 
in turning a given $(M,g)$ into a ``standard model''. This is 
almost always expressed in terms of different coordinate systems
on (subsets of) $M$, which hides the fact that a cosmological 
model can, of course, be fully characterised by proper 
geometric structures, independent of any choices of 
coordinates. 

\subsection{Standard Models}
The so-called standard models in relativistic cosmology are based 
on metrics of the FLRW (Friedman, Lema\^{\i}tre, Robertson,
Walker) form 
\begin{equation}
\label{eq:FLRW-Metrics}
g=c^2\,dt^2-a^2(t)g^{(3)}_{\rm cc}\,,
\end{equation}
where $g^{(3)}$ is a 3-dimensional Riemannian metric of constant 
curvature (hence subscript cc). A set of minimal assumptions 
leading to \eqref{eq:FLRW-Metrics} are given 
in~\cite{Straumann:1974}. The topologies of the constant time 
hypersurfaces $t=\mathrm{const.}$ are restricted if one requires  
for them the condition of completeness (equivalent to geodesic 
completeness). This implies closure (compact without boundary) 
and hence finite volume in the positive curvature case, whereas 
in the zero or negative curvature cases open and closed universes 
may exist.\footnote{Sometimes constant negative curvature is 
taken synonymously for openness, which is unfortunate.}

The geodesic vector field $V$ is given in these coordinates simply 
by 
\begin{equation}
\label{eq:GeodVFinFLRW}
V=\frac{\partial}{\partial t}\,.
\end{equation}
It is Killing if and only if $\dot a=0$. In the cases we 
are interested in here $a$ will have a non trivial time 
dependence, which means that neighbouring geodesics change 
separation. For infinitesimally nearby geodesics the 
perpendicular connecting vector, $X$,  changes according to 
Jacobi's equation 
\begin{equation}
\label{eq:JacobiEq}
\nabla_V\nabla_VX=-R(X,V)V\,,
\end{equation}
where $R$ is the Riemann tensor.\footnote{Its definition 
is $R(X,Y)Z:=\nabla_X\nabla_YZ-\nabla_Y\nabla_XZ-\nabla_{[X,Y]}Z.$}
For given $V$ the right-hand side of \eqref{eq:JacobiEq} should 
be read as a map $X\mapsto -R(X,V)V$ that at each point $p$ in 
spacetime linearly maps the 3-dimensional orthogonal complement 
of $V$ in tangent space to itself. It is called the Jacobi map 
$J$. For \eqref{eq:FLRW-Metrics} one finds  
\begin{equation}
\label{eq:JacobiMapFLRW}
J=\frac{\ddot a}{a}\Identity
\end{equation}
where $\Identity$ refers to the identity map (in the respective 
orthogonal complement of $V$). 

This is the general-relativistic rationale for the heuristic step
taken in \eqref{eq:ModNewEq-1}: If we write down \eqref{eq:JacobiEq}
in terms of Fermi normal coordinates $(t, \vec x)$ centred 
around some integral curve of $V$ (along which the eigentime 
equals cosmological time $t$), it follows immediately 
from~\eqref{eq:JacobiMapFLRW} that the nearby geodesics obey 
\begin{equation}
\label{eq:ObserverEqMot}
\ddot{\vec x}-(\ddot a/a)\vec x=0\,,
\end{equation}
where an overdot stands for differentiation with respect to $t$.
This equation characterises inertial motions in a small 
neighbourhood of a reference observer.  It can be generalised 
to the motion of electric charges in a FLRW background (i.e. 
without taking into account any back-reaction of the charges 
onto the spacetime geometry) in the following manner: 
Let a charge $Q$ move on an integral line of $V$ (i.e. 
geodesically). Determine its electromagnetic field by solving 
Maxwell's equations in that background.%
\footnote{This is easy since Maxwell's equations are 
conformally invariant and \eqref{eq:FLRW-Metrics} is 
conformally static (pull out $a^2(t)$ and introduce 
``conformal time'' $\eta$ via $d\eta=dt/a(t)$).} 
Consider another charge, $e$, that moves in the combined 
backgrounds of the FLRW universe and the electromagnetic 
field of the other charge. The law of motion for $e$ is that 
where the Lorentz four-force divided by the rest mass $m$ 
of $e$ replaces the zero on the right hand side of the geodesic 
equation. In a slow-motion approximation, where terms quadratic
and higher in $v/c$ are neglected (here $v$ refers to the 
velocity of $e$ relative to the notion of rest defined by $V$), 
the result takes the form~\cite{Carrera.Giulini:2010a}
\begin{equation}
\label{eq:LorentzForcedGeodesic}
\ddot{\vec x} -\frac{\ddot a}{a}\,\vec x=
\frac{eQ}{4\pi\varepsilon_0m}\cdot\frac{\vec x}{\Vert\vec x\Vert^3}\,,
\end{equation}
where a dot denotes derivatives with respect to cosmic time $t$ 
(proper time along integral curves of $V$) and $\vec x:=(x^1,x^2,x^3)$
are the Fermi proper-length coordinates in the surfaces of 
constant cosmic time. As far as this Hydrogen-atom-type situation
is concerned, the upshot is that we may indeed just take 
the familiar flat-space non-relativistic equation and replace  
$\ddot{\vec x}$ with $\ddot{\vec x} -\frac{\ddot a}{a}\,\vec x$,
as heuristically anticipated in the pseudo Newtonian picture 
of section\,\ref{sec:PseudoNewtionanPic}. 

Since the gravitational interaction of two bodies works via their  
back-reaction onto the geometry of space time, it is clear that for 
the gravitational analogue of \eqref{eq:LorentzForcedGeodesic} we 
cannot just work in the FLRW background. Even under the simplifying 
assumption that one mass can be treated as test particle (no back 
reaction) we still need to consider solutions to Einstein's equation
representing a central mass in an expanding universe. This will be 
discussed in section\,\ref{sec:BH-Expansion}. But before that 
we will have to say a few more words on the apparently simple 
equation~\eqref{eq:ObserverEqMot}.

\subsection{Mapping out trajectories}  
Equation \eqref{eq:ObserverEqMot} characterises an inertial structure
in a neighbourhood of a reference observer (who moves himself on a 
geodesic). This is achieved through the characterisation of inertial 
trajectories (i.e. characterising a path structure~\cite{Ehlers.Koehler:1977})
in particular coordinates, here Fermi normal coordinates. In this 
section we merely wish to emphasise the dependence of this procedure 
of  ``mapping out'' nearby trajectories on the simultaneity structure
employed, which should be made explicit in order to avoid confusion. 

With reference to a single selected observer $O$ (i.e. worldline; 
here even a geodesic), mapping out a nearby worldline $W$ usually 
means to establish a coordinate system $K$ in a tubular neighbourhood
$U$ of $O$ so that $W$ is contained in $U$. Such a system may, e.g.,  
consist of a parametrisation of $O$ in terms of eigentime and a 
foliation of $U$ by spacelike hypersurfaces intersecting $O$ orthogonally. 
A point $q$ is then given the following coordinates of \emph{time} and 
\emph{distance} (we ignore the angular part here): Let $p$ be the unique 
point on $O$ that lies on the same leaf $L$ of our foliation as $q$. 
Then the time coordinate of $q$ is the eigentime of $p$. 
Further, the distance of $q$ is the $L$-geodesic distances between 
$p$ and $q$. Here ``$L$-geodesic'' means that the curve is running 
entirely in $L$ and is a geodesic with respect to the metric that $L$
inherits from the ambient spacetime. 

Fermi normal coordinates are of that type and exist in any spacetime. 
Here the 3-dimensional leafs of the spacelike foliation are 
spanned by the geodesics emanating from, and perpendicular to, $O$.
This obviously implies that each leaf has zero extrinsic curvature at 
its intersection point with $O$. 

On the other hand, in an FLRW spacetime, we do have the natural foliation 
by hypersurfaces of homogeneity, i.e. constant cosmological time $t$. 
Here $t$ coincides with the eigentime for the (geodesic) reference 
observer. However, the cosmological foliation is certainly \emph{not} 
the same as that resulting from Fermi coordinates. This can immediately 
be seen from the fact that the extrinsic curvature of a leaf of 
constant cosmological time is proportional to its intrinsic metric
(hence the hypersurface is totally umbilical) and the Hubble constant 
as constant of proportionality. In particular, it never 
vanishes.

One should therefore not expect \eqref{eq:ObserverEqMot} to 
be of exactly the same analytic form as the equation one gets 
from the exact geodesic equation in FLRW spacetime in the 
standard FLRW coordinates. Let us be explicit here: A geodesic 
in a spatially flat (just for simplicity) FLRW universe is 
most easily obtained from the variational principle 
\begin{equation}
\label{eq:FLRW-Geodesics-1}
\delta\,\int d\tau\
\frac{1}{2}
\left[c^2\left(\frac{dt}{d\tau}\right)^2
-a^2(t)\left(\frac{d\vec y}{d\tau}\right)^2
\right]\,.
\end{equation}
Its Euler Lagrange equations are
\begin{equation}
\label{eq:FLRW-Geodesics-2}
\begin{split}
c^2&\frac{d^2t}{d\tau^2}+
a\dot a\left(\frac{d\vec y}{d\tau}\right)^2=0\,,\\
&\frac{d^2\vec y}{d\tau^2}\,+\,
2\frac{\dot a}{a}\frac{dt}{d\tau}\frac{d\vec y}{d\tau}\,=0\,.
\end{split}
\end{equation}
In order to compare this with \eqref{eq:ObserverEqMot} we need 
to rewrite this in a twofold way. The first is to replace $\tau$ with
$t$ as a parameter. The equation for $\vec y(t)$ then becomes
\begin{equation}
\label{eq:FLRW-Geodesics-3}
\ddot{\vec y}+2\frac{\dot a}{a}\dot{\vec y}
-\dot aa\,\frac{\Vert\dot{\vec y}\Vert^2}{c^2}\,\dot{\vec y}=0\,.
\end{equation}
The second step is to use instead of $\vec y$ the spatial
coordinate $\vec x:=a\vec y$, which has direct metric relevance
since the modulus of $\vec x$ gives the geodesic distance 
within the surface of constant cosmological time, unlike the 
``co-moving'' coordinate $\vec y$, the modulus of which corresponds 
to a distance proportional to $a(t)$. In terms of $\vec x$
equation \eqref{eq:FLRW-Geodesics-3} reads
\begin{equation}
\label{eq:FLRW-Geodesics-4}
\ddot{\vec x}-(\ddot a/a)\vec x=
\frac{\Vert\dot{\vec x}-(\dot a/a)\vec x\Vert^2}{c^2}\
\frac{\dot a}{a}\
\left(\dot{\vec x}-\frac{\dot a}{a}\vec x\right)\,.
\end{equation}
This equation is exact. It differs from \eqref{eq:ObserverEqMot}
by the terms on the right-hand side, which are small of 
order $v^2/c^2$. But one should still keep in mind that 
\eqref{eq:FLRW-Geodesics-3} refers to cosmological 
simultaneity whereas in \eqref{eq:ObserverEqMot} we used 
simultaneity of Fermi normal coordinates. 

In many applications within our solar system  so called 
\emph{radar coordinates} are used, which define yet another 
relation of simultaneity. These coordinates are again based 
on an observer $O$ and exist in a tubular neighbourhood $U$. 
We take $O$ to be parametrised by eigentime $\tau$ and use it 
to assign a time and distance value to any event $q$ in $U$
as follows: Let $p_+(q)$ and $p_-(q)$ be the intersections of 
the future and past light-cone at $q$ with $O$ respectively 
and $\tau_\pm(q)$ the corresponding parameter values. Then 
\begin{equation}
\label{eq:RadarCoord}
T(q)=\tfrac{1}{2}\bigl(\tau_+(q)+\tau_-(q)\bigr)\,,\quad
R(q)=\tfrac{c}{2}\bigl(\tau_+(q)-\tau_-(q)\bigr)
\end{equation}
are the time and distance assigned to $q$. Event $q$ is 
then simultaneous with event $p$ on $O$ half way 
between $p_+$ and $p_-$, i.e. the eigentime at $p$ is 
the arithmetic mean $\tfrac{1}{2}(\tau_++\tau_-)$. 
The hypersurface $T=T(q)$ intersects $O$ at $p$ 
perpendicularly. The ``radius'' $R(q)$  is generally 
different from the hypersurface geodesic distance between 
$p$ and $q$ (unlike the other cases discussed above).
This follows already from the simple observation that the 
coordinate functions $T$ and $R$ are invariant under all 
those conformal transformations $g\mapsto\phi^2 g$ of the 
metric for which $\phi\big\vert_O=1$, whereas this is certainly 
not true for Fermi normal coordinates.     

\begin{figure}
\begin{center}
\includegraphics[width=0.4\linewidth]{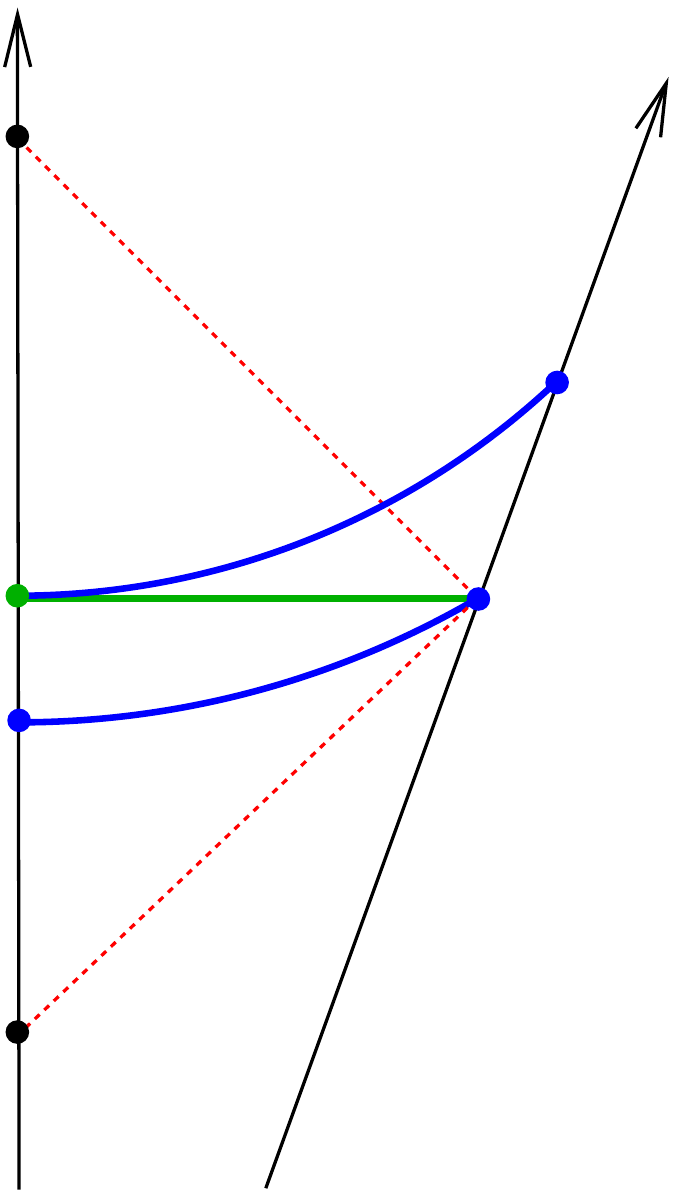}
\put(-145,257){\large $O$}
\put(-2,242){\large $W$}
\put(-36,124){\large $q$}
\put(-155,125){\large $p$}
\put(-155,99){\large $p'$}
\put(-160,223){\large $p_+$}
\put(-160,33){\large $p_-$}
\put(-138,111){\small $\Delta\tau$}
\end{center}
\caption{\label{fig:FigSimultaneity}\footnotesize
Radar versus cosmological simultaneity.
See main text for explanation.} 
\end{figure}
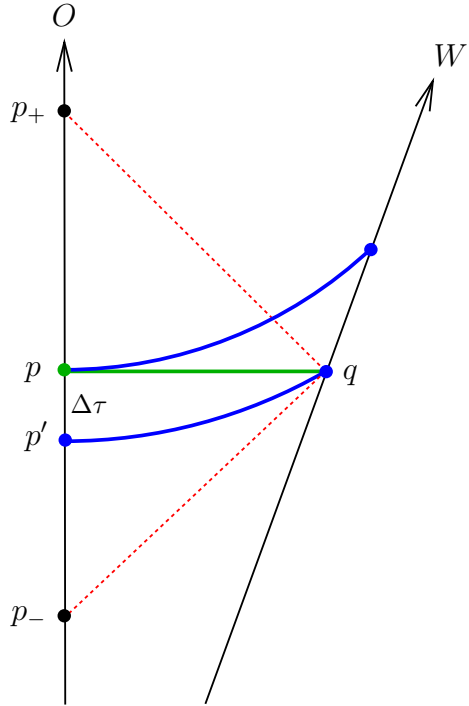 
Figure\,\ref{fig:FigSimultaneity} shows two worldlines: the 
observer $O$ and another one, $W$, corresponding to a spacecraft, 
say. The future and past light cones (dashed lines) of the 
event $q$ on $W$ intersect $O$ at $p_+$ and $p_-$ respectively. 
This can be interpreted as a light signal being emitted by the 
observer at $p_-$, reflected by a spacecraft moving at $q$, and 
received back by the observer at event $p_+$. In radar coordinates 
the event $p$ on the observer's worldline $O$ that is simultaneous 
with $q$ is that half way (in terms of eigentime along $O$) 
between $p_-$ and $p_+$. In contrast, the hypersurfaces 
$t=\mathrm{const.}$ of cosmological simultaneity correspond 
to the horizontal upward-bent curves. The lower one is that 
intersecting the reflection event $q$. Its intersection with 
the observer's worldline is at $p$ an eigentime $\Delta\tau$ 
prior to $p$. 

Assigning a distance to event $q$ on $W$ with respect to $O$ 
can either mean to assign the geodesic distance along the 
curved line $p'q$ to the event $p'$ on $O$, or to assign 
$R(q)$ as in \eqref{eq:RadarCoord} to the event $p$ on $O$. 
Depending on which one is used the distance of the 
spacecraft as function of proper time along $O$ is given by 
different functions. Consequently, this is also true for the 
relative velocity and acceleration. The following relations 
have been shown to hold in leading order 
\cite{Carrera.Giulini:2010a}:
\begin{subequations}
\label{eq:TaylorExpOrbitAll}
\begin{alignat}{2}
\label{eq:TaylorExpOrbit3a}
  &\tilde r &&= r-(H_0c)\ \tfrac{1}{2}(v/c)(r/c)^2\\
\label{eq:TaylorExpOrbit3b}
  &\tilde v &&= v-(H_0c)\ (v/c)^2(r/c)\\
\label{eq:TaylorExpOrbit3c}
  &\tilde a &&= a-(H_0c)\ \bigl\{(v/c)^3+(r/c)(v/c)(a/c)\bigl\} \,. 
\end{alignat}
\end{subequations}
Here $(r,v,a)$ are the distance, velocity, and acceleration
with respect to cosmological simultaneity and proper geodesic 
distance, whereas the quantities with tildes refer to radar 
coordinates. $H_0$ is the Hubble constant. Note in particular 
the first term on the right-hand side of \eqref{eq:TaylorExpOrbit3c} 
which shows an apparent additional inward pointing acceleration 
proportional to $H_0c$ in radar coordinates as compared to 
cosmological coordinates. Recall also that the Pioneer 
spacecrafts were tracked by Doppler methods so that indeed 
$\tilde a$ rather than $a$ was measured. However, as seen from 
\eqref{eq:TaylorExpOrbit3c}, the additional term $\propto H_0c$
is multiplied with $(v/c)^3$, which for the Pioneer spacecraft is 
of the order of $10^{-12}$. Hence this additional acceleration,
albeit proportional to $H_0c$, is strongly suppressed by the 
third power of $v/c$. Finally we mention that cosmic expansion 
also affects the method of Doppler tracking \emph{per se}. 
A detailed study of this has been performed 
in~\cite{Carrera.Giulini:2006}.

\section{Black holes and expansion} 
\label{sec:BH-Expansion} 
Intuitively we expect the local geometric properties of 
black holes to be affected if the black hole is placed into 
a cosmological environment. Anticipated changes could concern
the mass, the horizon structure, and certainly the orbits
of bound systems. More generally, one might fear that the
very notion of a ``black hole'' does not generalise in an 
obvious way. But before going into some of these aspects, 
we must mention that the very meaning of ``being placed'' 
is unclear in view of the fact that solutions to Einsteins 
equations cannot be simply superposed. Hence it is not obvious 
at first how to meaningfully compare an ordinary black hole 
solution corresponding to an asymptotically Minkowskian 
spacetime to an inhomogeneous cosmological solution that 
asymptotically approaches a FLRW universe and contains 
an inner event (or apparent) horizon. There is no natural 
notion of ``sameness'' by which we could identify black-hole 
solutions with different asymptotics. A somewhat pragmatic 
way, that wish to follow here, is to use (quasi-)local 
geometric properties as characterisations. One could then 
ask for the effect of cosmic expansion on the relation of 
such local geometric features, like, e.g., the change of 
horizon size for given mass. But for this  to make sense 
``horizon size'' and ``mass'' must first be defined 
geometrically. This can be done at least in the spherically 
symmetric case. To see this, we need to recall a few 
mathematical facts.  

\subsection{Geometric background}
\label{sec:GeomBackground}
We are used to the fact that standard cosmological spacetimes 
have a preferred foliation by spacelike hypersurfaces of 
constant cosmic time. In fact, the hypersurfaces are defined 
geometrically as the orbits of the symmetry group of spatial 
homogeneity and isotropy, and ``cosmic time'' is essentially 
the parameter that labels these hypersurfaces. In this 
subsection we wish to point out that the weaker condition of
spherical symmetry (without homogeneity) in fact suffices 
to geometrically characterise a foliation by spacelike 
hypersurfaces. Roughly speaking, the hypersurfaces are the 
spacelike hypersurfaces within which the areas of the 
2-spheres traced out by the action of the rotation group (its 
orbits) increase or decrease fastest, as compared to all 
other spatial directions perpendicular to the orbits themselves.

In order to say this more precisely, we first recall that 
a spacetime $(M,g)$ is called spherically symmetric 
if and only if there is an action of the rotation group $SO(3)$ on 
$(M,g)$ by isometries, such that the orbits of this 
action are spacelike 2-spheres. $M$ is then foliated 
by these 2-spheres, which means that each point $p$ 
in $M$ lies on precisely one such sphere. Let $a(p)$ be  
the (2-dimensional) volume of that sphere (i.e. its 
``surface area''), as measured by the spacetime metric 
$g$. Then we can define the following function 
\begin{equation}
\label{eq:DefArealRadius}
R: M\mapsto\reals_+\,,\qquad
R(p):=\sqrt{\frac{a(p)}{4\pi}}\,.
\end{equation}
This function is called the \emph{areal radius}. It assigns 
a positively-valued ``radius'' to each $SO(3)$ orbit, such 
that the 2-dimensional volume (surface area) of this orbit
is $4\pi R^2$, just as in ordinary flat space. Note that 
it is not proper to think of this radius as some kind of 
(geodesic-)distance to an ``origin'', since such an origin
would correspond to a fixed point of the $SO(3)$ action 
which we excluded here explicitly. It may not exist at all, 
even though the manifold is inextensible (i.e. no point
has been artificially removed), like e.g. in the maximally 
extended Schwarzschild-Kruskal spacetime. 

The function $R$ is $SO(3)$ invariant by construction. 
Hence the exterior differential, $dR$, which is a co-vector 
field on $M$, is also $SO(3)$ invariant.  We assume,
or restrict attention to that part of spacetime where 
this is true, that $dR$ is spacelike, so that $R$ is a 
good spatial radial coordinate and that the hypersurfaces 
of constant $R$ are timelike. This gives us a first and 
rather obvious foliation of spacetime by hypersurfaces. 
But there is another one: At each point in spacetime there 
is a unique line (unoriented direction) perpendicular to 
the $SO(3)$ orbit which is also annihilated by $dR$. 
Up to overall sign it can be represented by a normalised 
vector field $k$ orthogonal to the $SO(3)$ orbits and 
satisfying $dR(k)=0$, meaning that $R$ does not change 
along the flow of $k$. Now, $k$ is always hypersurface 
orthogonal (this statement only depends only on the line 
field represented by $k$). Hence we have a foliation of 
$M$ by spacelike hypersurfaces $\Sigma$ orthogonal 
to $k$, so that a leaf $R=\mathrm{const.}$ of the 
former foliation intersects a leaf $\Sigma$ of the 
latter in precisely one $SO(3)$ orbit. In this way,
the geometric structure imposed by spherical symmetry 
leads to a \emph{specific} foliation of spacetime 
into spacelike hypersurfaces, each leaf of which is 
itself foliated by $SO(3)$ orbits.\footnote{%
That spherically symmetric spacetimes admit spacelike 
foliations each leaf of which is foliated by $SO(3)$ 
orbits is not at all surprising: There are zillions 
of it. The point made here is, that there is, in a 
sense, a natural one.}
If the latter are parametrised by spherical polar 
coordinates $(\theta,\varphi)$, then $(R,\theta,\varphi)$ 
parametrise each spacelike leaf $\Sigma$. The most 
general spherically symmetric spacetime metric is 
then of the form
\begin{equation}
\label{eq:SphSymmSpacetimeMetric}
g=A^2(T,R)c^2\,dT^2-B^2(T,R)\,dR^2
-R^2\bigl(d\theta^2+\sin^2\theta\,d\varphi^2\bigr)\,.
\end{equation}
where $A,B$ are some non-vanishing dimensionless 
functions. 

\subsection{Reissner-Nordstr\"om-de\,Sitter}
For example, a spherically symmetric solution to the 
coupled Einstein-Maxwell system with cosmological 
constant $\Lambda$ (and no sources for the Maxwell 
field) is the Reissner-Nordstr\"om-de\,Sitter solution, 
where  
\begin{equation}
\label{eq:RNdeSitter}
A^2=B^{-2}=1-\frac{2m}{R}+\frac{q}{R^2}-\frac{\Lambda}{3}R^2\,.
\end{equation}
The parameters $m$ and $q$ have physical dimension of 
length (geometric units). They can be converted to 
parameters $M$ and $Q$ with physical dimensions of 
mass and electric charge in MKSA-units via
\begin{equation}
\label{eq:PhysicalMassCharge}
m=\frac{GM}{c^2}\,,\qquad
q^2=\frac{1}{4\pi\varepsilon_0}\frac{GQ^2}{c^4}\,.
\end{equation}
Here we continue to work in geometric units. 
The identification of $q$ as electric charge is 
unproblematic because $q$ can be shown to be the 
flux of the electric field through any of the 
2-spheres of constant $R$ (and hence any 2-surface
homologous to it). The identification of $m$ as 
mass is less obvious due to the lack of an 
unanimously accepted definition of (quasi-)local 
mass in GR. It is true that 
\eqref{eq:PhysicalMassCharge} becomes the 
Reissner-Nordstr\"om solution for $\Lambda=0$,
which is asymptotically flat and has $M$ as its 
well defined overall mass (i.e. at spacelike infinity) 
But for general $\Lambda$ \eqref{eq:RNdeSitter} is 
not asymptotically flat and we cannot just continue 
to call the parameter $m$ its mass without further 
justification. 

\subsection{Misner-Sharp mass and its refinement}
Fortunately, for spherically symmetric situations there 
is a reasonable notion of mass associated with each 2-sphere 
$R=\mathrm{const.}$ that is also easy to work with. 
Based on \cite{Misner.Sharp:1964}
it is called the Misner-Sharp mass (or energy). 
It has been shown to have a number of physically 
appealing properties \cite{Hayward:1996} and agrees with 
the more generally defined Hawking mass~\cite{Hawking:1968}
whenever the Misner-Sharp mass can be defined. 
A geometric definition can be given as follows~\cite{Carrera.Giulini:2010a}:
\begin{equation}
\label{eq:MisnerSharpGeom}
\MS(S)=-\frac{R^3}{2}\,\Sec(S)\,,
\end{equation}
where $\Sec(S)$ denotes the sectional curvature of spacetime $(M,g)$ 
tangent to the sphere $S$ of constant $R$.\footnote{%
Note that $\Sec(S)$ is not the intrinsic curvature 
of $S$.} of constant $R$. The overall minus-sign 
on the right hand side of \eqref{eq:MisnerSharpGeom} 
is due to our ``mostly-minus'' signature convention.   
Note that $\MS$ can be considered as function on 
the spacetime manifold $M$. Its value $\MS(p)$ at 
the point $p$ is simply $\MS(S)$, where $S$ is 
the unique 2-sphere ($SO(3)$ orbit) through $p$. 
It may then be shown that~\cite{Carrera.Giulini:2010a} 
\begin{equation}
\label{eq:MisnerSharpAlg}
\MS=\frac{R}{2}\Bigl(1+g^{-1}(dR,dR)\Bigr)\,.
\end{equation}
The result for metrics of the form 
\eqref{eq:SphSymmSpacetimeMetric} is immediate:
\begin{equation}
\label{eq:MisnerSharpEval1}
\MS(T,R)=\frac{R}{2}\Bigl(1-B^{-2}(T,R)\Bigr)\,.
\end{equation}
Further specialised to \eqref{eq:RNdeSitter} we get 
\begin{equation}
\label{eq:MisnerSharpEval2}
\MS(R)=m-\frac{q^2}{2R}+\frac{\Lambda}{6}R^3\,.
\end{equation}
The interpretation of the second and third term on the 
right-hand side is rather obvious. First, $q^2/2R$ is, 
in geometric units, the electrostatic field energy 
stored in that part of space where the areal radius 
is larger than $R$. This can be easily verified 
by direct computation from the explicit form of 
the electromagnetic field and its energy-momentum 
tensor. Without $\Lambda$, $m$ would be the total 
mass of the hole, \emph{including} its electrostatic field. 
$\MS(R)$ is then its total mass minus the electromagnetic 
energy located outside $S$. Second, the $\Lambda$ term in 
Einstein's equations corresponds to a mass 
density, which in geometric units is  
\begin{equation}
\label{eq:LambdaMassDensity}
\rho_\Lambda
=\frac{G}{c^2}\cdot\frac{1}{c^2}\cdot\frac{c^4}{8\pi G}\Lambda
=\frac{\Lambda}{8\pi}\,.
\end{equation}
Hence the last term on the right-hand side of 
\eqref{eq:MisnerSharpEval2} equals 
$\rho_\Lambda(4\pi/3)R^3$. This is a familiar 
formula in GR: a mass density 
multiplied by $(4\pi/3)R^3$ gives the total 
mass within radius $R$ \emph{diminished by the 
gravitational binding energy}. The last fact is 
hidden by this deceptively simple formula, but 
note that $(4\pi /3)R^3$ is generally \emph{not} 
the geometric volume enclosed by the sphere 
of areal radius $R$.

Given \eqref{eq:MisnerSharpEval2} we may ask how to separate 
the black hole mass from the other components due to the 
electric field and the cosmological constant \emph{in a geometric way}. 
It has been argued in \cite{Carrera.Giulini:2010a}
quite generally that this can be achieved by splitting the 
sectional curvature in \eqref{eq:MisnerSharpGeom} into its 
Weyl and its Ricci part. Indeed, in the special case at hand 
the first term on the right-hand side of 
\eqref{eq:MisnerSharpEval2} (the $m$) is the Weyl part, 
the other two terms comprise the Ricci part. This also works in 
other spherically-symmetric situations, as we shall 
discuss below. 

\subsection{Applications}
\subsubsection*{Reissner--Nordstr\"om--de\,Sitter spacetimes}
We now have the tools at hand to discuss relations between 
geometrically defined quantities of spherically symmetric 
black holes in different 
environments. If we identify the mass of the black hole with 
the Weyl part of the Misner-Sharp mass, we may meaningfully 
compare the areal radii of its horizon for fixed mass with 
or without $\Lambda$. In the present example 
\eqref{eq:RNdeSitter} this just boils down to 
discussing the dependence on $\Lambda$ of the 
root of $A^2(R)$. The root corresponds to the hole's 
horizon (event or apparent, as we are in a static situation). 

For $\Lambda=0$ a horizon exists if $m\geq\vert q\vert$ 
and lies at an areal radius
\begin{equation}
\label{eq:HorizonRadius}
R_{\rm Hor}=m+\sqrt{m^2-q^2}\,.
\end{equation} 
(The ``inner'' root at $m-\sqrt{m^2-q^2}$ corresponds to a Cauchy 
horizon and does not interest us here.) Switching on $\Lambda$ 
shifts this root to 
\begin{subequations}
\label{eq:HorizonRadiusRNDS}
\begin{equation}
\label{eq:HorizonRadiusRNDS-a}
R_{\rm Hor}\rightarrow \tilde 
R_{\rm Hor}:=R_{\rm Hor}(1+\epsilon)\,,
\end{equation}
where, to for small $\Lambda R_H^2$, we get in leading order
\begin{equation}
\label{eq:HorizonRadiusRNDS-b}
\varepsilon=\frac{\Lambda}{6}\,R_{\rm Hor}^3\,\frac{1}{\sqrt{m^2-q^2}}\,.
\end{equation}
\end{subequations}
Hence $\tilde R_{\rm Hor}$ is larger/smaller than $R_{\rm Hor}$ if $\Lambda$
is larger/smaller than zero. This might be taken to conform 
with ones intuition that an accelerated/decelerated expansion 
somehow "`pulls/pushes"' the horizon to larger/smaller radii.
But this is again deceptive since the horizon is not a material 
substratum acted upon by forces. Since $u:=A^{-1}\partial/\partial T$ 
is the four velocity of the static observer, his acceleration 
measured in his instantaneous rest frame is 
\begin{equation}
\label{eq:AccelerationStaticObs}
a=\nabla_uu=\frac{c^2}{2}\,B\,\frac{dA^2}{dR}\,e_1=
c^2\frac{
\frac{m}{R^2}-\frac{q^2}{R^3}-\frac{\Lambda R}{3}}%
{\sqrt{1-\frac{2m}{R}+\frac{q^2}{R^2}-\frac{\Lambda}{3}R^2}}\,
e_1\,,
\end{equation}
where $e_1:=B^{-1}\partial/\partial R$ is the normal vector 
in radial direction. Hence for $\Lambda$ switched on the 
acceleration already diverges at a radius larger than 
$R_{\rm Hor}$ because of its effect on the relevant zero of 
the expression under the root in the denominator 
(which is $A^2$), not because 
of its effect in the numerator. Recall also that the 
extreme case occurs for such $(m,q,\Lambda)$ where the 
two hole horizons coincide. For $\Lambda=0$ this happens 
for $\vert q\vert=m^2$, but for $\Lambda>0$ at values  
$\vert q\vert>m$. This last fact generalises to the 
rotation parameter in the Kerr--de\,Sitter family of 
rotating holes, as was recently discussed in some detail 
in~\cite{Akcay.Matzner:2011}.

\subsubsection*{Einstein-Straus vacuoles}
Another elegant application of the Misner-Sharp mass is 
to derive the radius of the vacuole in the Einstein-Straus
model~\cite{Einstein.Straus:1945}. This model, also called the 
``Swiss-Cheese model'' consists of matching a 
Schwarzschild-de\,Sitter (or Kottler \cite{Kottler:1918})
solution to a FLRW universe along some radius, along which 
the outward pull from the cosmological masses are just balanced 
by the inward pull from the central black hole. The matching 
surface in spacetime is a timelike hypersurface foliated by 
the $SO(3)$ orbits. Each such orbit represents the matching 
surface at a time. We wish to determine its radius. 

The matching conditions are traditionally given 
by the continuity of the induced metrics (first fundamental 
forms) and extrinsic curvatures (second fundamental forms) 
defined on either sides of the hypersurface, the so-called 
Lanczos-Darmois-Israel conditions~%
\cite{Lanczos:1924}\cite{Darmois:1927}\cite{Israel:1966,Israel:1967}.
Another but equivalent set of conditions involves the areal 
radii and Misner-Sharp energies, stating their equality for 
each pair of $SO(3)$ orbits to be 
matched~\cite{Carrera:2009,Carrera.Giulini:2010a}.%
\footnote{The partial statement, that the (in our terminology) 
Misner-Sharp masses of the excised ball and the inserted 
inhomogeneity have to be equal in order for the resulting 
spacetime to satisfy Einstein's equation, is also known 
as \emph{Eisenstaedt's Theorem}~\cite{Eisenstaedt:1977}.}

The Schwarzschild--de\,Sitter metric is given by 
\eqref{eq:SphSymmSpacetimeMetric}\eqref{eq:RNdeSitter}, 
where $q=0$. The FLRW cosmological model is given by the metric  
\begin{equation}
\label{eq:FLRW-Metric}
g=c^2\,dt^2-a^2(t)\left(\frac{dr^2}{1-kr^2}
+r^2\bigl(d\theta^2+\sin^2\theta\,d\varphi^2\bigr)\right)\,,
\end{equation}
and the four-velocity field $V=\partial/\partial t$.
For an energy-momentum tensor of the form \eqref{eq:DustEMT} 
(perfect fluid) Einstein's equations are then 
equivalent to Friedmann's equations
\begin{subequations}
\label{eq:FriedmannEquations}
\begin{alignat}{2}
\label{eq:FriedmannEquations-a}
&\frac{{\dot a}^2}{a^2}&&\,=\,
 \frac{\Lambda c^2}{3}
+\frac{8\pi G}{3}\,\rho
-\frac{kc^2}{a^2}\,,\\
\label{eq:FriedmannEquations-b}
&\frac{\ddot a}{a}&&\,=\,
\frac{\Lambda c^2}{3}
-\frac{4\pi G}{3}\left(\rho+\frac{3p}{c^2}\right)\,.
\end{alignat}
\end{subequations}
From \eqref{eq:FLRW-Metric} we immediately read of the 
areal radius $R(t,r)=a(t)r$, so that $dR=\dot ar\,dt+a\,dr$.
Inserting this into the expression \eqref{eq:MisnerSharpAlg}
for the Misner-Sharp mass gives 
\begin{equation}
\label{eq:MS-Energy-FLRW}
\MS(R)
=\tfrac{1}{2}Rr^2\bigl(k+c^{-2}{\dot a}^2\bigr)
=\frac{4\pi}{3}R^3\,\bigl[\rho_{\rm matter}+\rho_\Lambda\bigr]\,,
\end{equation}
where $\rho_{\rm matter}$ and $\rho_\Lambda$ are the mass 
densities of the dust and cosmological constant, respectively, 
in geometric units, that is $\rho_{\rm matter}=\rho G/c^2$ and 
$\rho_\Lambda$ as in \eqref{eq:LambdaMassDensity}. In the second 
step in~\eqref{eq:MS-Energy-FLRW} we used the fist Friedmann
equation \eqref{eq:FriedmannEquations-a} to eliminate 
${\dot a}^2$. Note that neither the pressure $p$ of 
the matter nor the curvature $k$ of space enters the 
final expression in \eqref{eq:MS-Energy-FLRW}.

In \eqref{eq:MisnerSharpEval2} we already calculated the 
Misner-Sharp mass for the Reissner-Nordstr\"om--de\,Sitter 
case. Setting $q=0$ we obtain the Misner-Sharp mass for the 
Schwarzschild--de\,Sitter case. It contains two terms, the 
second being equal to the second in \eqref{eq:MS-Energy-FLRW}; 
compare \eqref{eq:LambdaMassDensity} and line below. 
Here we already used that the matching conditions require the 
areal radii of the matching spheres to be equal. Hence, in 
order for the Misner-Sharp masses of the matching spheres 
in the FLRW universe and the Schwarzschild--de\,Sitter 
universe to be equal, the first terms must also coincide. 
This immediately gives the simple condition
\begin{subequations}
\label{eq:Schuecking}
\begin{equation}
\label{eq:SchueckingMass}
m=\frac{4\pi}{3}R^3\rho_{\rm matter}
\end{equation}
or, equivalently,
\begin{equation}
\label{eq:SchueckingRadius}
R=\left(\frac{3m}{4\pi\rho_{\rm matter}}\right)^{1/3}\,.
\end{equation}  
\end{subequations}
Again note that this expression makes no explicit 
reference to the pressure (and hence no reference to 
an equation of state for the matter) and the curvature.

If space were flat \eqref{eq:SchueckingMass} just said that the 
mass concentrated in the black hole just equals that fraction 
that formerly had been homogeneously distributed inside the 
excised ball (the vacuole), and \eqref{eq:SchueckingRadius} said 
that in order to include a black hole of mass $m$ into 
a FLRW universe one has to remove the cosmological mass 
surrounding it up to that radius inside which the cosmological 
dust has an integrated mass of just $m$. In the general (curved) 
cases \eqref{eq:Schuecking} is almost deceptively simple, since 
for spaces of constant positive/negative curvature the expression 
$(4\pi/3)R^3$ grows slower/faster with the areal radius $R$ 
than the actual geometric volume.\footnote{The first correction 
for the volume inside a 2-sphere of areal radius $R$ in a 
3-space of constant cuvature $k=\pm 1$ is 
$\mathrm{Vol}_k(R)=(4\pi/3)R^3(1+3kR^2/10+...$.}

\subsubsection*{Mc\,Vittie--type spacetimes}
Finally I wish to mention attempts to interpolate 
between spacetime metrics representing black holes 
in a near zone and FLRW cosmologies in a far zone.
Most of what follows will be based on 
\cite{Carrera.Giulini:2010a} and \cite{Carrera.Giulini:2010b}.

First attempts in this direction go back to McVittie in 
the early 1930s~\cite{McVittie:1933}, who attempted 
a particle-like interpretation. The basic idea is to 
literally interpolate analytically between two metrics.
Since the result of a na\"{\i}ve interpolation will generally 
depend on the coordinates used, it is necessary 
to do this in a geometrically meaningful way. 
We recall that any spherically symmetric spatial 
metric is conformally flat. So the first thing to do 
is to write the black-hole metric as well as the 
FLRW metric in a manifest spatially conformally flat 
form, like 
\begin{equation}
\label{eq:SchwarzschildIsotrCoord}
g_{\rm BH}
=\left[\frac{1-\frac{m}{2r}}{1+\frac{m}{2r}}\right]^2\,c^2dt^2-
\left[1+\frac{m}{2r}\right]^4
\bigl(dr^2+r^2(d\theta^2+\sin^2\theta\,d\varphi^2)\bigr)
\end{equation}
and
\begin{equation}
\label{eq:FLRWConfFlat}
g_{\rm Cosm}
=c^2\,dt^2-a^2(t)
\bigl(dr^2+r^2(d\theta^2+\sin^2\theta\,d\varphi^2)\bigr)\,.
\end{equation}
Expression \eqref{eq:SchwarzschildIsotrCoord} corresponds
to \eqref{eq:SphSymmSpacetimeMetric} with $q=\Lambda=0$ and 
after redefinition of the radial coordinate. Expression 
\eqref{eq:FLRWConfFlat} is just \eqref{eq:FLRW-Metric} 
with $k=0$ set for simplicity, so that the metric 
is already in manifest conformally flat form. 
The interpolation now consists in writing the McVittie form 
(hence subscript MV)
\begin{equation}
\label{eq:McVittieAnsatz}
g_{\rm MV}
=\left[\frac{1-\frac{m(t)}{2r}}{1+\frac{m(t)}{2r}}\right]^2\,c^2dt^2-
a^2(t)\left[1+\frac{m(t)}{2r}\right]^4
\bigl(dr^2+r^2(d\theta^2+\sin^2\theta\,d\varphi^2)\bigr)\,.
\end{equation}
Note that formally this just changes 
\eqref{eq:SchwarzschildIsotrCoord} by 1)~writing a 
$a^2(t)$ in front of the spatial part and 2)~allowing 
$m$ to become time dependent. The general strategy is 
now to evaluate the left-hand side of Einstein's equations 
using \eqref{eq:McVittieAnsatz}, and then ``see'' - basically 
by trial and error - what can reasonably be put on the 
right hand side.\footnote{This may be called the ``poor 
man's way to solve Einstein's equations''.} 

The relaxation of allowing $m$ to be time dependent seems 
clearly necessary if we wish to discuss processes like 
the accretion of cosmological matter by the black hole through 
radial infall.\footnote{Clearly it has to be radial since we 
restricted to spherical symmetric situations.} However,
the change from \eqref{eq:SchwarzschildIsotrCoord} to 
\eqref{eq:McVittieAnsatz} will also bring about a change 
in the very notion of mass. Here it becomes important that we 
can distinguish between the mass of the central object and 
that of cosmological matter (forming overdensities, say). 
This can be done using the refinement of the Misner-Sharp 
mass discussed above~\cite{Carrera.Giulini:2010a}. 

An obvious way to proceed is to just read off the areal radius 
from \eqref{eq:McVittieAnsatz}: 
\begin{equation}
\label{eq:McVittieArealRadius}
R(t,r)=\left[1+\frac{m(t)}{2r}\right]^2\,a(t)r
\end{equation}
and then calculate the Misner-Sharp according to 
\eqref{eq:MisnerSharpAlg}. If we denote by $e_0$ 
the timelike unit vector normal to the hypersurfaces
$t=\mathrm{const.}$, i.e. parallel to $\partial/\partial t$,
and by $e_1$ the spacelike unit vector in radial direction,
i.e. parallel to $\partial/\partial r$, we can rewrite
\eqref{eq:MisnerSharpAlg} as\footnote{%
We recall that, because of spherical symmetry, 
the hypersurfaces $t=$ const. and the spatial radial 
directions (being tangential to these hypersurfaces and 
normal to the $SO(3)$ orbits in them) can be characterised 
purely geometrically, as explained in 
section\,\ref{sec:GeomBackground}.} 
\begin{equation}
\label{eq:MS-MassOnMcVittie}
\MS=
\frac{R}{2}\Bigl(1-\bigl[e_1(R)\bigr]^2\Bigr)+ 
\frac{R}{2}\bigl[e_0(R)\bigr]^2\,.
\end{equation}
It can be shown \cite{Carrera.Giulini:2010a} that the first term is 
the Weyl part of the Misner-Sharp mass, whereas the second part is 
its Ricci part. Straightforward evaluation of the first part using 
\eqref{eq:McVittieArealRadius} gives $ma$, whereas the second term 
can be shown to be related to the $(e_0,e_0)$-component of the 
Einstein Tensor. It total we get  
\begin{equation}
\label{eq:McVittieMS-Mass-1}
\MS=am+\tfrac{1}{6}R^3\mathbf{Ein}(e_0,e_0)
=\MS^{\rm Weyl}+\MS^{\rm Ricci}\,.
\end{equation}
We can use Einstein's equation to replace the Einstein 
tensor in the second term on the right-hand side with the 
energy momentum tensor of matter and the cosmological constant. 
A possible cosmological constant adds a term $R^3\Lambda/6$, 
just like in \eqref{eq:MisnerSharpEval2}, which again may 
be interpreted as additional homogeneous mass density 
given by \eqref{eq:LambdaMassDensity}. If we assume the 
energy-momentum tensor to be that of a perfect 
fluid \eqref{eq:DustEMT} moving along trajectories 
orthogonal to the hypersurfaces $t=$ const., that is 
\begin{equation}
\label{eq:NoInfluxCond}
V=ce_0\,,
\end{equation}
the $(e_0,e_1)$-component of Einstein's equation implies 
\begin{equation}
\label{eq:NoMassAccretion}
m(t)a(t)=:m_0=\mathrm{const.}
\end{equation}
Equation \eqref{eq:McVittieMS-Mass-1} then takes the form 
\begin{equation}
\label{eq:McVittieMS-Mass-2}
\MS=m_0+
\frac{4\pi}{3}R^3\,\bigl[\rho_{\rm matter}+\rho_\Lambda\bigr]\,,
\end{equation}
in which the second term is just that which we already 
obtained in \eqref{eq:MS-Energy-FLRW} for the pure FLRW 
case and which here, as before, appears as the Ricci 
part of the Misner-Sharp mass. Its Weyl part turns out 
to be a constant $m_0$ given by $ma$ (not just $m$!).
Again we stress that in general $(4\pi/3)R^3$ is not 
the space volume of a ball or radius $R$ so that the 
right-hand side of \eqref{eq:McVittieMS-Mass-2} is \emph{not} 
the space integral over the mass density; the difference 
being due to the gravitating binding energy. 

Clearly, the constancy of the Weyl part, which we defined 
with the mass of the inhomogeneity, is a consequence 
of our assumption \eqref{eq:NoInfluxCond}, which means that 
there is no accretion of cosmological matter by the central 
object. This ``object'' need not be a black hole. An interior 
solution modelled on a homogeneous perfect-fluid ``star''
of positive mass density is known \cite{Nolan:1992}. 
The four-velocity of the star's surface is $V$, i.e. it is 
co-moving, so that \eqref{eq:NoInfluxCond} indeed is a 
condition for no-accretion. Note that ``co-moving'' means 
constant $r$, which according to \eqref{eq:McVittieArealRadius}
implies an expanding areal radius $R$, essentially proportional 
to $a(t)$ when $m_0\ll ar$. 

It is clear that the no-accretion condition can only be 
achieved through a special pressure function that just 
balances gravitational attraction. It turns out that this 
can be relaxed but, somewhat surprisingly, only at the 
price of introducing heat flows. This means that 
\eqref{eq:DustEMT} needs to be generalised to 
\begin{equation}
\label{eq:PerfectFluidFrictionEMT}
T=\rho V\otimes V +p\bigl(c^{-2}V\otimes V-g^{-1})
+c^{-2}\bigl(Q\otimes V+V\otimes Q\bigr)\,,
\end{equation}
where $Q$ is a spacelike vector perpendicular to $V$ 
that represents the residual current-density of energy 
(heat) in the rest system of the fluid. The matter is 
now not moving along $e_0$, but rather has a 
non-vanishing radial velocity relative to the 
frame $(e_0,e_1)$. Expressed in terms of the rapidity 
$\chi$ one has 
\begin{subequations}
\label{eq:InfallMatterAndHeat}
\begin{alignat}{2}
\label{eq:InfallMatter}
V=c\bigl(\cosh\chi\,e_0+\sinh\chi\,e_1\bigr)\,,\\
\label{eq:InfallHeat}
Q=q\bigl(\sinh\chi\,e_0+\cosh\chi\,e_1\bigr)\,.
\end{alignat}
\end{subequations}
Now, equations~\eqref{eq:PerfectFluidFrictionEMT} and 
\eqref{eq:InfallMatterAndHeat} allow for solutions of
Einstein's equation in the form \eqref{eq:McVittieAnsatz} 
only if~\cite{Carrera.Giulini:2010b}
\begin{equation}
\label{eq:Constraint}
(c^2\rho+p)\tanh\chi+2q/c=0\,.
\end{equation}
This constraint shows that radial infall of matter $\chi<0$
must be accompanied by a radial outflow of heat $q>0$, and 
vice versa. For pressures $p$ small compared to $\rho c^2$ 
the modulus of the infalling matter-energy exceeds that 
of the outflowing heat by a factor of two.%
\footnote{Note that the rapidity $\chi$ is 
related to the ordinary velocity by $\tanh\chi=v/c$.} 
This means that infalling matter always increases the mass 
$\MS^{\rm Weyl}$ at a rate roughly half of the net infall of 
matter energy divided by $c^2$. (This is just what the 
$(0,1)$-component of Einstein's equation expresses.)
But the increase of inertial mass implies an increase in 
gravitational mass. A slow-motion and weak-field
approximation of the geodesic equation in 
\eqref{eq:McVittieAnsatz}, which gives it a pseudo-Newtonian 
form, shows that the Newtonian potential is proportional 
to $\MS^{\rm Weyl}/R$, so that the gravitational pull
increases with increasing mass. As a result, orbits of 
test particles will spiral ``inwards'', i.e. to smaller 
$R$-values~\cite{Carrera.Giulini:2010a,Carrera.Giulini:2010b}.

Note that at the level of our discussion \eqref{eq:Constraint}
is of purely geometric origin and does not lie in any deeper 
physical insight into heat production by infalling matter due 
to friction, as we had not specified any such model. 
The geometric origin is inherent in the Ansatz 
\eqref{eq:McVittieAnsatz} and has been identified in \cite{Carrera.Giulini:2010b} as the condition of 
\emph{Ricci isotropy}, which means that the four-dimensional 
Ricci tensor and the four-dimensional spacetime metric are 
pointwise proportional after being pulled back to the 
hypersurfaces $t=$ const. This condition would have to 
be relaxed in order to specify radial flows of matter and 
heat independently, as seems physically desirable.

Finally we comment on the singularity structure of 
\eqref{eq:McVittieAnsatz}, following~\cite{Carrera.Giulini:2010b}.
As expected, it has a singularity at $r=0$ in the Weyl 
part of the curvature and, somewhat unexpected, a 
singularity in the Ricci part of the curvature at $r=m/2$. 
The latter is absent only in the following limiting cases: 
1)~$m=0$ and $a$ arbitrary (pure FLRW),
2)~$m$ and $a$ constant (pure Schwarzschild), and 
3)~$am=$ const. and $\dot a/a$=const. (Schwarzschild 
de\,Sitter case). The singularity at $r=m/2$ will lie within 
a trapped region, i.e. behind an apparent horizon. 
Since here the asymptotic structure of spacetime is 
considerably more difficult to analyse than in the 
previous cases, we restrict attention to the local 
concept of apparent horizons. Recall 
that a spacelike 2-sphere S is said to be trapped, marginally 
trapped, or untrapped if the product $\theta^+\theta^-$ of 
the expansions for the outgoing and ingoing future-pointing 
null vector fields normal to S is positive, zero, or negative,
respectively. An apparent horizon is the boundary of a trapped 
region.  

Now, in our case, the product $\theta^+\theta^-$ can be shown 
to equal
\begin{equation}
\label{eq:ProdExpansiona}
\theta^+\theta^-=\frac{g^{-1}\bigl(dR,dR\bigr)}{R^2}=\frac{2\MS-R}{R^3}\,,
\end{equation}
where the last equality following from \eqref{eq:MisnerSharpAlg}.
This shows that a trapped, marginally trapped, or untrapped region 
corresponds to $dR$ being timelike, lightlike, and spacelike 
respectively, or, equivalently, $2\MS-R$ being positive, zero, 
and negative respectively. 

According to the last equality in \eqref{eq:ProdExpansiona}
the location of the apparent horizons is given by the zeros 
of the function $2\MS-R$. In order to write this function 
in a convenient form, we restrict to expansion ($\dot a>0$) 
and introduce the following non-negative quantities:
\begin{equation}
\label{eq:ConvAbbreviations}
R_M:=2\MS^{\rm Weyl}=2ma\,,\quad
R_H:=\frac{c}{H_0}=\frac{ca}{\dot a}\,,\quad
x:=\frac{R}{R_M}\,.
\end{equation}
$R_M$ is the mass-radius just as in \eqref{eq:SchwarzschildRadius}, 
but now ``mass'' refers specifically to the mass of the central 
object that we identified with the Weyl part of the Misner-Sharp 
mass. $R_H$ is the Hubble radius defined in terms of the Hubble 
constant as before~\eqref{eq:HubbleRadius}. Finally it will be 
convenient to use the dimensionless variable $x$.
The last expression in \eqref{eq:ProdExpansiona}, divided by 
$R_M$, is then readily seen from \eqref{eq:McVittieMS-Mass-1} to be  
\begin{equation}
\label{eq:HorizonFunct-1}
F(x):=\frac{2\MS-R}{R_M}=1-x+\tfrac{1}{3}x^3 \ R^2_M\mathbf{Ein}(e_0,e_0)\,.
\end{equation}
The $(e_0,e_0)$-component of the Einstein tensor for the metric \eqref{eq:McVittieAnsatz} can be computed as function of $(t,r)$
and then re-expressed as function of $(t,x)$. The result is

\begin{equation}
\label{eq:HorizonFunct-2}
F(x):=\frac{2\MS-R}{R_M}=1-x+x^3\Bigl[\frac{R_M}{R_H}
+{\dot R}_M\,\Theta(x)\Bigr]^2\,,
\end{equation}
where $\Theta$ is some positive function which we need not 
specify here. Note the proportionality of the second term 
with ${\dot R}_M$, i.e. twice the time rate of change of the    
mass of the central black hole. It vanishes in the case of no 
mass accretion, in which case $F(x)$ just becomes a simple 
polynomial of third order which is already in reduced form 
(no $x^2$ terms). Its zeros can be written down explicitly 
using Cardano's formula, but the essential features can be 
seen directly. As $F(x)$ is positive for $x=0$ and 
$x\rightarrow\infty$, it has two zeros if and only if 
$F(x)$ assumes a negative value at its minimum, which is 
at $x=R_H/R_M\sqrt{3}$). This is the case if and only if 
$R_M<2R_H/3\sqrt{3}$, which is the case of interest here 
(the hole's horizon radius being much smaller than the Hubble radius).  
A leading order expansion for the  
location of the zeros of $F(x)$ for small values of 
$R_M/R_H$ is now simple. For $R_M/R_H=0$ we have $x=1$, 
i.e. $R_{\rm Hor}=R_M=2ma$. Switching on cosmological 
expansion shifts this root to 
\begin{subequations}
\label{eq:HorizonRadiusMcVittie}
\begin{equation}
\label{eq:HorizonRadiusMcVittie-a}
R_{\rm Hor}\rightarrow \tilde 
R_{\rm Hor}:=R_M(1+\epsilon)\,,
\end{equation}
where for small $R_M/R_H$ we now get to leading order
\begin{equation}
\label{eq:HorizonRadiusMcVittie-b}
\varepsilon=\left(\frac{R_M}{R_H}\right)^2\,.
\end{equation}
\end{subequations}
This is precisely the generalisation of \eqref{eq:HorizonRadiusRNDS} 
for $q=0$ and $\Lambda>0$, where $R_H=\sqrt{3/\Lambda}$ 
for de\,Sitter spacetime. In the Mc\,Vittie case, too, the 
areal radius of the apparent horizon is enlarged by 
expansion.
 
This discussion can be generalised to the case of non-zero 
mass accretion~\cite{Carrera.Giulini:2010b}. The result is 
that mass accretion further enlarges the (instantaneous) 
value of the apparent horizon's areal radius. 
This concludes our discussion of black holes in expanding 
universes. This is a difficult subject and not much is 
known in terms of exact solutions.  

\subsubsection*{Black-Hole cosmologies}
We have seen in some detail how a single black hole may inhibit
cosmological expansion locally. This was most pronounced in the 
Einstein-Strauss construction where the metric becomes static 
throughout the vacuoles of areal radius \eqref{eq:SchueckingRadius}.
As we have discussed at the end of 
section\,\ref{sec:Coulomb-like-potential}, a realistic size for 
such a vacuoles would be that of galaxy clusters. So it seems 
reasonable to approximate the dynamics of a closed universe above 
galaxy-cluster size by a finite number of vacuoles each with a 
galaxy-cluster mass at the centre. Could this, in turn, be 
approximated by the same number of black holes, each with 
galaxy-cluster mass, and without any other forms of matter? 
In this case Einstein's vacuum equations, possibly with 
cosmological constant, would suffice to discuss the dynamics 
of the universe, at least in this approximation.  
This idea has indeed been entertained by Lindquist \& Wheeler in 
a seminal paper in 1957~\cite{Lindquist.Wheeler:1957}. They consider 
a ``lattice universe'' made out of a finite number of black holes 
distributed on the surface of a 3-sphere according to the 
vertices of a convex regular polytope in four dimensional 
euclidean space (in which we think of the 3-sphere as being 
embedded such that the vertices of the polytope lie on it). 
Once the black holes are introduced the geometry changes of 
course and it is assumed, as noted by  Lindquist \& Wheeler, 
that ``This approximation demands that the distribution of
gravitational influences just external to each sphere should
depart relatively little from spherical symmetry''. This is 
modelled by the fact that \emph{regular} polytopes are chosen, 
which meets as close as possible the usual requirement of 
isotropy around each point (here vertex). In two space dimensions 
regularity means that the black holes are situated at the 
vertices of one of the five 
platonic solids inscribed into a 2-sphere. Such a spatially 
two-dimensional universe would consist of 4, 6, 8, 12, or 20 
black holes, corresponding to the vertices of the tetrahedron, 
octahedron, cube, icosahedron, or dodecahedron. In three spatial 
dimensions there exist five 
analogs of the platonic solids carrying 5, 8, 16, 120 and 600 
vertices, respectively, and one more with 24 vertices that has 
no direct analog.  The even simpler case of just two 
black holes sitting at antipodal points of the 3-sphere has not 
been considered by Lindquist \& Wheeler and will be discussed 
below. Initial data for the vacuum Einstein equations
corresponding to a given number of equal-mass black holes at given 
locations on a 3-sphere can be constructed by standard methods.
In the particular case of time symmetry (zero extrinsic curvature) 
and spatial conformal flatness such data can be written down 
explicitly; see, e.g., \cite{Brill.Lindquist:1963}, where 
Fig.\,3 shows the geometry of three black holes in a closed 
universe, and \cite{Giulini:1998a} for a general discussion.  
The time evolution according to Einstein's equations is a more 
complicated matter that cannot be dealt with without numerical 
integration. The quantity of interest is the distance between 
black-hole neighbours (suitably defined) as a function of time. 
This one may attempt for the simpler cases of the 5- and 8-hole 
universe~\cite{Lindquist.Wheeler:1957} and the qualitative 
behaviour is quite easily understood and does indeed resemble 
the standard FLRW dynamics; see, e.g.,\cite{Clifton.Ferreira:2009}.
The full field-theoretic problem remains of course a formidable 
task. 

Regarding the last point, we may try to get rid of the complicated 
dynamical issues by considering static situations, in the most simple case 
involving only two black-holes. To keep the holes at a constant 
distance we either have to introduce singularities on (segments of) 
the symmetry axis, or, more interestingly, a positive cosmological 
constant (positive energy density) whose negative pressure does the 
job. Such a solution has indeed already been envisaged by Erich Trefftz 
in 1922~\cite{Trefftz:1922}, who obtained just the Schwarzschild-de\,Sitter
solution (first found by Kottler \cite{Kottler:1918}) but endowed 
it with a different global interpretation, namely as representing 
\emph{two} Schwarzschild black holes located at antipodal points of a 3-sphere. Note that this solution still has the rotational $SO(3)$-symmetry. 
Einstein, in a critical reply to Trefftz paper (\cite{Einstein:CP13}, Doc.\,387, pp.\,595-596), observed from direct inspection of the metric written down by Trefftz that stationary points of the areal-radius 
function correspond to zeros of the metric coefficient in front of 
$dt^2$. As in Trefftz' solution the areal radius is not constant in 
the region between the black holes, Einstein concluded that it 
must assume a stationary point and hence that the time-time component 
of the metric must vanish somewhere. This Einstein (erroneously) 
interpreted as the indubitable sign of an additional singularity 
between the black holes, which would indeed render the interpretation 
given to it by Trefftz impossible. 

However, the proper geometric meaning of the vanishing of this 
particular metric coefficient is that the static Killing vector 
field becomes lightlike. Physically it means that the 
acceleration of the static observers approaching this critical 
set diverge. This indicates a horizon rather than a singularity.
(See \cite{Geyer:1980} and \cite{UzanEllisLarena:2011} for a 
discussion of the global properties of the Schwarzschild-de\,Sitter 
solution in terms of Penrose diagrams.) 
Hence we arrive at the following question: Are there solutions to Einstein's 
equations with cosmological constant representing two spherically 
symmetric stars at constant distance without horizons in the 
region between the stars? This question has recently been 
addressed and answered in the negative for black holes in 
\cite{UzanEllisLarena:2011} and for perfect fluid stars 
in \cite{BoehmerFodor:2008}. However, the non-existence 
result for black holes presented in \cite{UzanEllisLarena:2011} 
depends on an assumption (non-constancy of the areal radius) 
which may be dropped so that the actual set of solutions is 
larger than anticipated. 

Einstein's mathematical observation regarding the connection between stationary points of the areal-radius function and the location of 
horizons was made on the basis of  Trefftz' formulae for the 
Schwarzschild-De\,Sitter solution. But it is not hard to see that it 
is really of a more general kind. This is already implicit in equation 
\eqref{eq:ProdExpansiona}, which informs us that zeros of $dR$ 
correspond to marginally trapped regions.  Let us therefore see what can be said on the basis of Einstein's equations alone. We are interested in 
static and spherically symmetric solutions to Einstein's vacuum 
equations with cosmological constant in which the areal radius 
might assume stationary points or be even piecewise constant. 
Hence we write the metric in the form 
\begin{equation}
\label{eq:Nariai-1}
g=f^2(r)\,c^2dt^2-dr^2
-R^2(r)\bigl(d\theta^2+\sin^2\theta\,d\varphi^2\bigr)\,.  
\end{equation}
We have chosen the radial coordinate such that $g_{rr}=-1$. 
The areal radius $R$ may then be a general function of $r$,
i.e. not restricted to $dR\ne 0$, as it would be if $R$ were 
taken to be a coordinate. Einstein's vacuum equations with 
cosmological constant are then equivalent to the following 
set of three equations, corresponding to the $tt$, $rr$, and 
$\theta\theta$ (equivalently $\varphi\varphi$) components 
respectively:
\begin{subequations}
\label{eq:Nariai-EinstEq}
\begin{alignat}{2}
\label{eq:Nariai-EinstEq-a}
&\Lambda+f''/f+2f'R'/fR&&=0\,,\\
\label{eq:Nariai-EinstEq-b}
&\Lambda+f''/f+2R''/R  &&=0\,,\\
\label{eq:Nariai-EinstEq-c}
&\Lambda+R''/R+f'R'/fR-R^{-2}(1-R'^2)&&=0\,.
\end{alignat}
\end{subequations}
Taking the difference between \eqref{eq:Nariai-EinstEq-a} 
and  \eqref{eq:Nariai-EinstEq-a}  gives 
\begin{equation}
\label{eq:Nariai-EinstEq-Lemma}
fR''=f'R'\,.
\end{equation}
Now suppose $r=r_*$ is a stationary point for $R$, i.e. $R'(r_*)=0$.
Then \eqref{eq:Nariai-EinstEq-Lemma} shows that $f(r_*)=0$ if
$R''(r_*)\ne 0$, i.e. if $R$ assumes a proper extremal value at 
$r_*$. But zeros of $f$ correspond to horizons, which is 
Einstein's observation (in modern terminology and interpretation). 
But we can also see that there is precisely one way to avoid this 
argument, namely if $R$ is constant; $R=R_0$. Equation \eqref{eq:Nariai-EinstEq-c} then gives
\begin{equation}
\label{eq:Nariai-Radius}
R_0=1/\sqrt{\Lambda}\,,
\end{equation}
whereas equations \eqref{eq:Nariai-EinstEq-a} and 
\eqref{eq:Nariai-EinstEq-b} both give $f''=-\Lambda f$
(harmonic-oscillator equation). The two integration constants 
(amplitude and phase) can be absorbed by redefining the 
scale of the $t$ and the origin of the $r$ coordinate. 
This leads to the \emph{Nariai} metric (in static form),
known since 1950 \cite{Nariai:1999-a}\cite{Nariai:1999-b}:
\begin{equation}
\label{eq:Nariai-Metric}
g=\cos^2\bigl(r/R_0\bigr)\,c^2dt^2-dr^2-
R_0^2(d\theta^2+\sin^2\theta\,d\varphi^2)\,.
\end{equation}
The possibility to connect two black holes via a piece of the 
Nariai metric, which is not considered in \cite{UzanEllisLarena:2011}, is presently investigated~\cite{Fennen:2013}. 
Some solutions where the Nariai metric connects two perfect-fluid stars were discussed in \cite{BoehmerFodor:2008}.

\subsection*{Acknowledgements}
This is a written-up version of a talk delivered at the conference 
on \emph{Philosophical Aspects of Modern Cosmology} held in Granada, 
22-23 September 2011. I am very grateful to Henrik Zinkernagel 
for inviting me to this most stimulating conference at this 
fantastic place!

\bibliographystyle{plain}
\bibliography{RELATIVITY,HIST-PHIL-SCI,MATH,QM,COSMOLOGY}
\end{document}